\PassOptionsToPackage{unicode}{hyperref}
\PassOptionsToPackage{hyphens}{url}
\PassOptionsToPackage{dvipsnames,svgnames,x11names}{xcolor}
\documentclass[
]{article}

\usepackage{amsmath,amssymb}
\usepackage{iftex}
\ifPDFTeX
  \usepackage[T1]{fontenc}
  \usepackage[utf8]{inputenc}
  \usepackage{textcomp} 
\else 
  \usepackage{unicode-math}
  \defaultfontfeatures{Scale=MatchLowercase}
  \defaultfontfeatures[\rmfamily]{Ligatures=TeX,Scale=1}
\fi
\usepackage{lmodern}
\ifPDFTeX\else  
\fi
\IfFileExists{upquote.sty}{\usepackage{upquote}}{}
\IfFileExists{microtype.sty}{
  \usepackage[]{microtype}
  \UseMicrotypeSet[protrusion]{basicmath} 
}{}
\makeatletter
\@ifundefined{KOMAClassName}{
  \IfFileExists{parskip.sty}{%
    \usepackage{parskip}
  }{
    \setlength{\parindent}{0pt}
    \setlength{\parskip}{6pt plus 2pt minus 1pt}}
}{
  \KOMAoptions{parskip=half}}
\makeatother
\usepackage{xcolor}
\makeatletter
\ifx\paragraph\undefined\else
  \let\oldparagraph\paragraph
  \renewcommand{\paragraph}{
    \@ifstar
      \xxxParagraphStar
      \xxxParagraphNoStar
  }
  \newcommand{\xxxParagraphStar}[1]{\oldparagraph*{#1}\mbox{}}
  \newcommand{\xxxParagraphNoStar}[1]{\oldparagraph{#1}\mbox{}}
\fi
\ifx\subparagraph\undefined\else
  \let\oldsubparagraph\subparagraph
  \renewcommand{\subparagraph}{
    \@ifstar
      \xxxSubParagraphStar
      \xxxSubParagraphNoStar
  }
  \newcommand{\xxxSubParagraphStar}[1]{\oldsubparagraph*{#1}\mbox{}}
  \newcommand{\xxxSubParagraphNoStar}[1]{\oldsubparagraph{#1}\mbox{}}
\fi
\makeatother

\providecommand{\tightlist}{%
  \setlength{\itemsep}{0pt}\setlength{\parskip}{0pt}}\usepackage{longtable,booktabs,array}
\usepackage{calc} 
\usepackage{etoolbox}
\makeatletter
\patchcmd\longtable{\par}{\if@noskipsec\mbox{}\fi\par}{}{}
\makeatother
\IfFileExists{footnotehyper.sty}{\usepackage{footnotehyper}}{\usepackage{footnote}}
\makesavenoteenv{longtable}
\usepackage{graphicx}
\makeatletter
\newsavebox\pandoc@box
\newcommand*\pandocbounded[1]{
  \sbox\pandoc@box{#1}%
  \Gscale@div\@tempa{\textheight}{\dimexpr\ht\pandoc@box+\dp\pandoc@box\relax}%
  \Gscale@div\@tempb{\linewidth}{\wd\pandoc@box}%
  \ifdim\@tempb\p@<\@tempa\p@\let\@tempa\@tempb\fi
  \ifdim\@tempa\p@<\p@\scalebox{\@tempa}{\usebox\pandoc@box}%
  \else\usebox{\pandoc@box}%
  \fi%
}
\def\fps@figure{htbp}
\makeatother
\NewDocumentCommand\citeproctext{}{}
\NewDocumentCommand\citeproc{mm}{%
  \begingroup\def\citeproctext{#2}\cite{#1}\endgroup}
\makeatletter
 \let\@cite@ofmt\@firstofone
 \def\@biblabel#1{}
 \def\@cite#1#2{{#1\if@tempswa , #2\fi}}
\makeatother
\newlength{\cslhangindent}
\newlength{\csllabelwidth}
\newenvironment{CSLReferences}[2] 
 {\begin{list}{}{%
  \setlength{\itemindent}{0pt}
  \setlength{\leftmargin}{0pt}
  \setlength{\parsep}{0pt}
  \ifodd #1
   \setlength{\leftmargin}{\cslhangindent}
   \setlength{\itemindent}{-1\cslhangindent}
  \fi
  \setlength{\itemsep}{#2\baselineskip}}}
 {\end{list}}
\usepackage{calc}

\usepackage{changepage}
\usepackage[font=footnotesize,labelfont=bf]{caption}
\makeatletter
\@ifpackageloaded{caption}{}{\usepackage{caption}}
\AtBeginDocument{%
\ifdefined\contentsname
  \renewcommand*\contentsname{Table of contents}
\else
  \newcommand\contentsname{Table of contents}
\fi
\ifdefined\listfigurename
  \renewcommand*\listfigurename{List of Figures}
\else
  \newcommand\listfigurename{List of Figures}
\fi
\ifdefined\listtablename
  \renewcommand*\listtablename{List of Tables}
\else
  \newcommand\listtablename{List of Tables}
\fi
\ifdefined\figurename
  \renewcommand*\figurename{Figure}
\else
  \newcommand\figurename{Figure}
\fi
\ifdefined\tablename
  \renewcommand*\tablename{Table}
\else
  \newcommand\tablename{Table}
\fi
}
\@ifpackageloaded{float}{}{\usepackage{float}}
\floatstyle{ruled}
\@ifundefined{c@chapter}{\newfloat{codelisting}{h}{lop}}{\newfloat{codelisting}{h}{lop}[chapter]}
\floatname{codelisting}{Listing}

\makeatother
\makeatletter
\@ifpackageloaded{caption}{}{\usepackage{caption}}
\@ifpackageloaded{subcaption}{}{\usepackage{subcaption}}
\makeatother

\usepackage{bookmark}

\IfFileExists{xurl.sty}{\usepackage{xurl}}{} 
\hypersetup{
  pdftitle={Big shells, bigger data: cohort analysis of Chesapeake Bay Crassostrea virginica reefs},
  pdfauthor={Madison D. Griffin\^{}\{1,*\}; Grace S. Chiu\^{}\{1,2,3,4,5\}; Roger L. Mann\^{}\{1\}; Melissa J. Southworth\^{}\{1\}; John K. Thomas\^{}\{1\}},
  colorlinks=true,
  linkcolor={blue},
  filecolor={Maroon},
  citecolor={Blue},
  urlcolor={Blue},
  pdfcreator={LaTeX via pandoc}}

\title{Big shells, bigger data: cohort analysis of Chesapeake Bay
\emph{Crassostrea virginica} reefs}
\author{Madison D. Griffin\(^{1,*}\) \and Grace S.
Chiu\(^{1,2,3,4,5}\) \and Roger L. Mann\(^{1}\) \and Melissa J.
Southworth\(^{1}\) \and John K. Thomas\(^{1}\)}
\date{}

\begin{document}
\maketitle

\small
$^1$William \& Mary's Batten School of Coastal \& Marine Sciences \& Virginia Institute of Marine Science

$^2$Computational \& Applied Mathematics \& Statistics (CAMS), College of William \& Mary

$^3$Department of Statistical Sciences and Operations Research, Virginia Commonwealth University

$^4$Department of Statistics and Actuarial Science, University of Waterloo

$^5$Department of Statistics, University of Washington

$^*$Corresponding author: mdgriffin@vims.edu
\normalsize
\vspace{1em}

\textbf{Abstract}

Oysters in Virginia Chesapeake Bay oyster reefs are ``age-truncated'',
possibly due to a combination of historical overfishing, disease
epizootics, environmental degradation, and climate change. Research has
suggested that oysters exhibit resilience to environmental stressors;
however, that evidence is based on the current limited understanding of
oyster lifespan. Until this paper, the Virginia Oyster Stock Assessment
and Replenishment Archive (VOSARA), a spatially and temporally expansive
dataset (222 reefs across 2003-2023) of shell lengths (SL, mm), had yet
to be examined comprehensively in the context of resilience. We develop
a novel method using Gaussian mixture modeling (GMM) to identify the age
groups in each reef using yearly SL data and then link those age groups
over time to identify cohorts and estimate their lifespan. Sixty-four
reefs (29\%) are deemed to have sufficient data (at least 300 oysters
sampled for a minimum of 8 consecutive years) for this analysis. We fit
univariate GMMs for each year (\(t\)) and reef (\(r\)) for each of the
seven river strata (\(R\)) to estimate 1) the mean and standard
deviation of SL for each \(a_{Rrt}\)th age group, and 2) the mixture
percentage of each \(a_{Rrt}\)th age group. We link age groups across
time to infer age cohorts by developing a mechanistic algorithm that
prevents the shrinking of shell length when an \(a_{Rrt}\)th group
becomes an (\(a_{R,r,t} + 1\))th group. Our method shows promise in
identifying oyster cohorts and estimating lifespan solely using SL data.
Our results show signals of resiliency in almost all river systems:
oyster cohorts live longer and grow larger in the mid-to-late 2010s
compared to the early 2000s.


\section{Introduction}\label{introduction}

\subsection{Historic declines in oyster
populations}\label{historic-declines-in-oyster-populations}

\begin{figure}[H]

\begin{minipage}{\linewidth}

\centering{

\pandocbounded{\includegraphics[keepaspectratio]{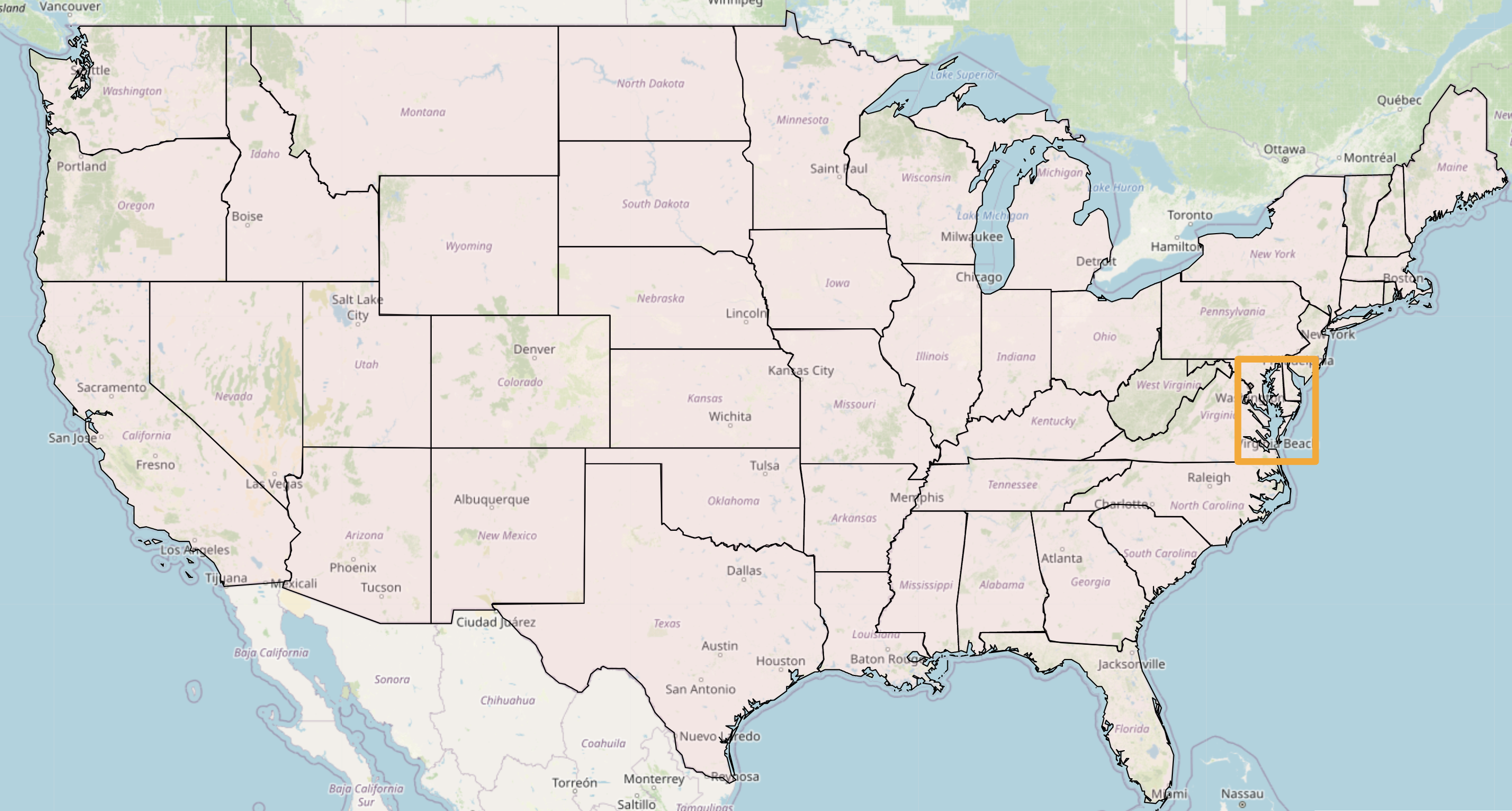}}

}

\subcaption{\label{fig-usmap}Map of the contiguous United States. The
Chesapeake Bay estuary is situated in the orange box.}

\end{minipage}%
\newline
\begin{minipage}{\linewidth}

\centering{

\pandocbounded{\includegraphics[keepaspectratio]{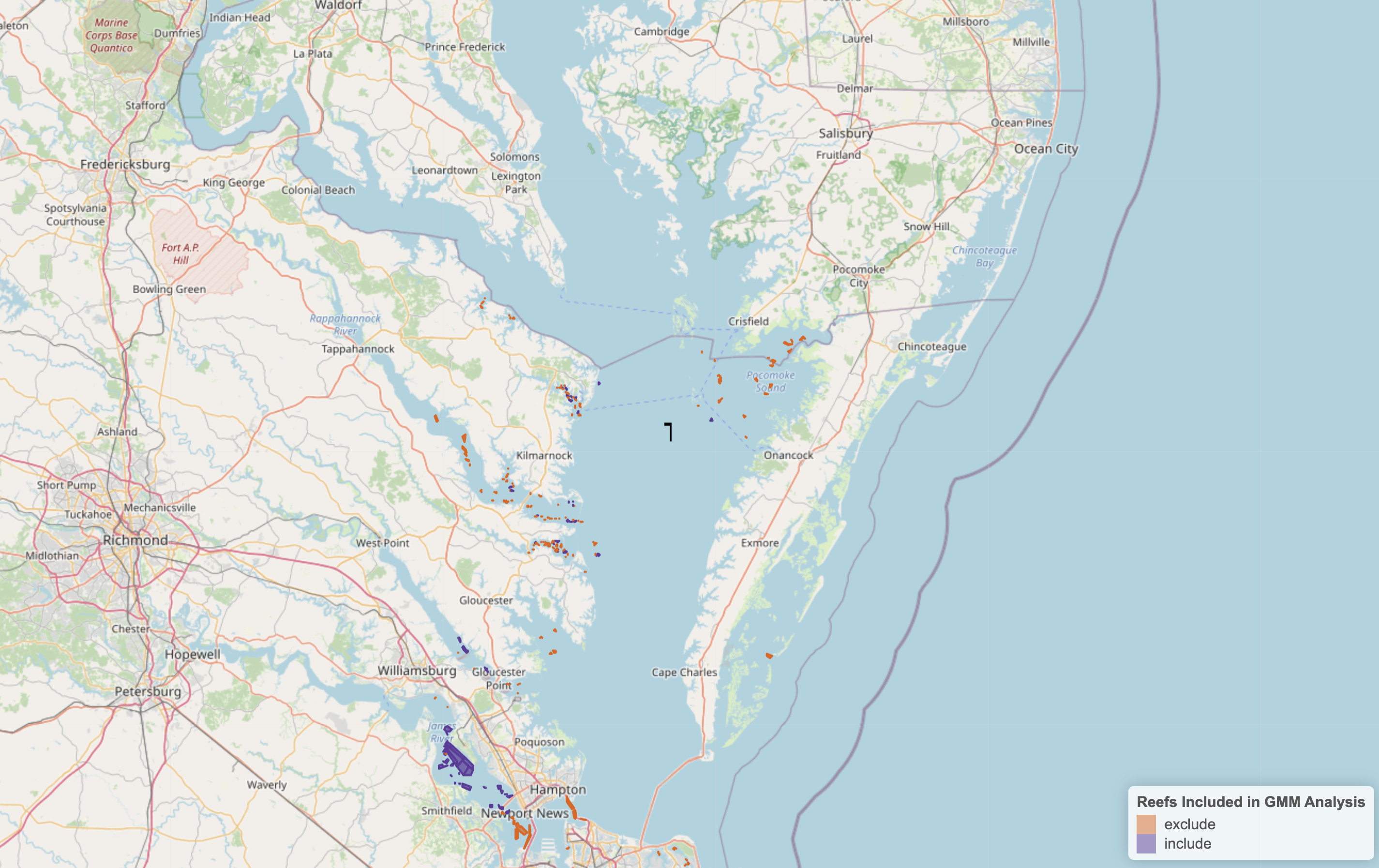}}

}

\subcaption{\label{fig-VACB}The map shows the reefs in the Virginia
portion of Chesapeake Bay (VACB) monitored by VOSARA. Reefs in purple (n
= 64) are included in the cohort analysis, meaning for at least 8
consecutive years there were at least 300 oysters sampled. Reefs
included in this study are located in every main VACB tributary except
Lynnhaven Bay.}

\end{minipage}%

\caption{\label{fig-mapbroad}A map of the study location and spatial
scale.}

\end{figure}

\phantom{xxxx}The eastern oyster (\emph{Crassostrea virginica}) has
played an important ecological, and cultural role in the Chesapeake Bay
in the eastern United States (Figure~\ref{fig-usmap}) and its estuaries
for over 10,000 years
(\citeproc{ref-mannPopulationStudiesNative2009}{Mann et al., 2009}). Our
study only considers the Virginia portion of Chesapeake Bay
(Figure~\ref{fig-VACB}). Eastern oyster reefs in the Chesapeake Bay
operate within a highly dynamic system, characterized by numerous daily,
weekly, and seasonal cycles, all of which are reflected in their shell
growth (\citeproc{ref-bairdSeasonalDynamicsChesapeake1989}{Baird \&
Ulanowicz, 1989}; \citeproc{ref-haasEffectSpringneapTidal1977}{Haas,
1977};
\citeproc{ref-pritchardSalinityDistributionCirculation1952}{Pritchard,
1952}). Oyster reefs are essential fish habitats, ecosystem engineers,
filter feeders, and on average, provide an estimated \$10,000 to
\$99,000 per hectare in value from ecosystem services
(\citeproc{ref-coenRoleOysterReefs1999}{Coen et al., 1999};
\citeproc{ref-grabowskiEconomicValuationEcosystem2012}{Grabowski et al.,
2012}; \citeproc{ref-paceDyingDecayingDissolving2020}{Pace et al.,
2020}). As ecosystem engineers, oyster reefs need to be self sustaining,
requiring oysters to live long, die at big sizes, and have ample shell
substrate for recruits to settle
(\citeproc{ref-hardingManagementPiankatankRiver2010}{Harding et al.,
2010}; \citeproc{ref-hemeonNovelShellStock2020}{Hemeon et al., 2020};
\citeproc{ref-paceDyingDecayingDissolving2020}{Pace et al., 2020};
\citeproc{ref-solingerOystersBegetShell2022}{Solinger et al., 2022}).

\phantom{xxxx}The Virginia oyster fishery peaked in the late 1800s;
however, fishery output varied from the 1880s through the 1910s
(\citeproc{ref-schulteHistoryVirginiaOyster2017}{Schulte, 2017}). The
public oyster replenishment program began in 1929 to address losses in
productivity (\citeproc{ref-schulteHistoryVirginiaOyster2017}{Schulte,
2017}). By the late 1940s, \emph{Perkinsus marinus}, a parasite known as
Dermo, caused annual mortality ranging from 10-30\% of market oysters
(\citeproc{ref-schulteHistoryVirginiaOyster2017}{Schulte, 2017}). In the
late 1950s, \emph{Haplosporidium nelsoni} (MSX) disease caused large
mortality events in Chesapeake Bay (90-95\%), which led to major
declines in oyster production
(\citeproc{ref-havenPublicOysterBottoms1986}{Haven \& Whitcomb, 1986};
\citeproc{ref-mannWhyOysterRestoration2007}{Mann \& Powell, 2007};
\citeproc{ref-schulteHistoryVirginiaOyster2017}{Schulte, 2017}). Both
diseases are still a prominent issue today. This series of epizootics
depressed the fishery in the 1980s through 1990s
(\citeproc{ref-schulteHistoryVirginiaOyster2017}{Schulte, 2017}).
Conservative management and sustained replenishment efforts by the
Virginia Marine Resource Commission resulted in gradual gains through
the early 2000's with a sustained improvement in stocks noted in the
2006-2010 period, and continuing growth since that time.

\phantom{xxxx}Oyster populations are successful at high densities due to
the life history strategy of accumulating into three dimensional reefs
(\citeproc{ref-hemeonNovelShellStock2020}{Hemeon et al., 2020};
\citeproc{ref-schulteHistoryVirginiaOyster2017}{Schulte, 2017}). As
such, oyster reefs require ample suitable oyster bottom and shell to aid
in recruitment and post-settlement success
(\citeproc{ref-hemeonNovelShellStock2020}{Hemeon et al., 2020}). Oyster
death is what supplies shell to the reef
(\citeproc{ref-paceDyingDecayingDissolving2020}{Pace et al., 2020}).
When shell is lost, due to either fishing mortality or short-half life
of oyster shell, recruitment decreases, which decreases the amount of
adult oysters, further decreasing shell addition
(\citeproc{ref-paceDyingDecayingDissolving2020}{Pace et al., 2020};
\citeproc{ref-powellRiseFallCrassostrea2012}{Powell et al., 2012}). This
negative feedback loop is found in many oyster reefs in estuarine
environments, hence, net shell loss is greater than net shell gain
(\citeproc{ref-mannPopulationStudiesNative2009}{Mann et al., 2009};
\citeproc{ref-paceDyingDecayingDissolving2020}{Pace et al., 2020};
\citeproc{ref-powellHowLongDoes2006}{Powell et al., 2006};
\citeproc{ref-powellRiseFallCrassostrea2012}{Powell et al., 2012};
\citeproc{ref-southworthOysterCrassostreaVirginica2010}{Southworth et
al., 2010}). Therefore, habitat (\emph{i.e.,} other oysters, both living
and dead) are essential for reef sustainability and should be properly
managed (\citeproc{ref-mannWhyOysterRestoration2007}{Mann \& Powell,
2007}; \citeproc{ref-solingerOystersBegetShell2022}{Solinger et al.,
2022}). One method of doing so is to estimate the age structure
(\emph{i.e.,} identify age classes and estimate their age) of oyster
reefs, and simulate management reference points based on shell accretion
and age classes (\citeproc{ref-solingerOystersBegetShell2022}{Solinger
et al., 2022}). A reef with multiple age classes alive and growing
simultaneously are more suitable for rebuilding populations and
improving self-sustainability
(\citeproc{ref-mannWhyOysterRestoration2007}{Mann \& Powell, 2007}).

\phantom{xxxx}The longevity of recruited individuals is essential to
effectively rebuild native populations for the benefit of ecological
restoration (\citeproc{ref-mannWhyOysterRestoration2007}{Mann \& Powell,
2007}). Older oysters add more shell to a reef due to their bigger size,
which is crucial for maintaining reef structure
(\citeproc{ref-mannOysterShellProduction2022}{Mann et al., 2022};
\citeproc{ref-mannWhyOysterRestoration2007}{Mann \& Powell, 2007}).
However, research suggests that the Chesapeake Bay oyster population is
age-truncated, meaning that the proportion of older age classes is low
(\citeproc{ref-hardingManagementPiankatankRiver2010}{Harding et al.,
2010}; \citeproc{ref-mannPopulationStudiesNative2009}{Mann et al.,
2009};
\citeproc{ref-southworthOysterCrassostreaVirginica2010}{Southworth et
al., 2010}). Age-truncation has been apparent since the colonial time
period. In a study using size to approximate age, oysters were
significantly larger in the middle Pleistocene (maximum of 260 mm)
compared to colonial oysters (maximum of 124 mm), suggesting that even
colonial oysters, often considered the baseline state for the eastern
oyster population, were missing older age groups
(\citeproc{ref-kusnerikUsingFossilRecord2018}{Kusnerik et al., 2018}).
At the modern time scale, despite consistently high recruitment events,
high mortality between year 0 and year 1 continues to truncate
Chesapeake Bay's oyster population, hindering restoration and rebuilding
efforts (\citeproc{ref-mannPopulationStudiesNative2009}{Mann et al.,
2009}). For example, in the Piankatank and Great Wicomico Rivers, even
though recruitment was high, there were small percentages of age 3+
oysters leading up to the year 2010
(\citeproc{ref-hardingManagementPiankatankRiver2010}{Harding et al.,
2010};
\citeproc{ref-southworthOysterCrassostreaVirginica2010}{Southworth et
al., 2010}). Due to the age-truncation in Chesapeake Bay oysters, dating
back to the colonial period, widely used age-based models in fisheries
science (\emph{e.g.,} von Bertalanffy
(\citeproc{ref-vonbertalanffyQuantitativeTheoryOrganic1938}{1938})) are
ineffective, requiring innovative approaches using size data
(\citeproc{ref-mannOysterShellProduction2022}{Mann et al., 2022}).

\subsection{Statistical methods for aging
oysters}\label{statistical-methods-for-aging-oysters}

\phantom{xxxx}Researchers have been interested in estimating the age and
identifying multi-year cohorts of oysters in Chesapeake Bay; however,
few have used a rigorous modeling approach. In the 1954 Annual Report of
the Commission of Fisheries of Virginia
(\citeproc{ref-commissionoffisheriesofvirginiaFiftyforthFiftyfifthAnnual1954}{Commission
of Fisheries of Virginia, 1954}), the condition of public oyster grounds
was shown to have improved due to the tracking of spat to yearling
oysters in the James and Rappahannock rivers. However, this conclusion
was reached from simple visualizations of bar charts. By the 2010s,
research using the Virginia Oyster Stock Assessment and Replenishment
Archive (VOSARA) dataset
(\citeproc{ref-hardingManagementPiankatankRiver2010}{Harding et al.,
2010}; \citeproc{ref-mannPopulationStudiesNative2009}{Mann et al.,
2009};
\citeproc{ref-southworthOysterCrassostreaVirginica2010}{Southworth et
al., 2010}) used the Bhattacharya method
(\citeproc{ref-bhattacharyaSimpleMethodResolution1967}{Bhattacharya,
1967}) to estimate age classes and growth rate in the Piankatank, James,
and Great Wicomico Rivers. By producing frequency histograms of shell
lengths for each year, the Bhattacharya method identifies clusters of
shell lengths using cubic approximations of density for each histogram,
allowing the researchers to identify age classes by regarding each
cluster as an age class from that year. Though more rigorous than the
methods using in the 1950s, this method still relies more heavily on the
modeler's decision making rather than on statistical principles.
Further, studies in the 2010s were focused on one river system, rather
than holistic approaches to analyze age structure across Chesapeake Bay.
Mann et al. (\citeproc{ref-mannOysterShellProduction2022}{2022}) used
the \texttt{TropFishR} package
(\citeproc{ref-mildenbergerTropFishRPackageFisheries2017}{Mildenberger
et al., 2017}) to identify year classes, then applied Bhattacharya
methods to overlay a growth curve on the estimated year classes to
calculate mortality rates on a larger scale (regional aggregations of
Chesapeake Bay river systems in Virginia and Maryland).

\phantom{xxxx}Alongside the ecological research community, the
statistical community has worked to improve methods to estimate age
groups from length data
(\citeproc{ref-macdonaldAgeGroupsSizeFrequencyData1979}{Macdonald \&
Pitcher, 1979}). Knowing that this is a common research goal in
fisheries science, Macdonald \& Pitcher
(\citeproc{ref-macdonaldAgeGroupsSizeFrequencyData1979}{1979}) developed
a computer program that transitioned the field from graphical
interpretations to model-based approaches by fitting a ``distribution
mixture''. This method builds off the work of Bhattacharya
(\citeproc{ref-bhattacharyaSimpleMethodResolution1967}{1967}), improving
reproducibility by reducing the amount of intepretation left to the user
(which can be biased)
(\citeproc{ref-macdonaldAgeGroupsSizeFrequencyData1979}{Macdonald \&
Pitcher, 1979}). One such approach is Gaussian mixture modeling (GMM).
This method has been used in ecological studies
(\citeproc{ref-fabri-ruizBenthicEcoregionalizationBased2020}{Fabri-Ruiz
et al., 2020}; \citeproc{ref-weisePedigreeAnalysisEstimates2022}{Weise
et al., 2022}) and fisheries science to estimate age groups
(\citeproc{ref-laslettFittingGrowthModels2004}{Laslett et al., 2004};
\citeproc{ref-sethiAccurateAgingJuvenile2017}{Sethi et al., 2017};
\citeproc{ref-shawCohortAnalysisEuphausia2021}{Shaw et al., 2021};
\citeproc{ref-zhouBayesianHierarchicalApproach2020}{Zhou et al., 2020})
based on animal size measurements.

\phantom{xxxx}A mixture modeling approach has not been comprehensively
applied to the VOSARA dataset. We propose a novel algorithm that applies
GMM to the VOSARA dataset to effectively allocate individual oysters
(observed from each oyster reef taken from each river stratum) into one
of several age groups (Gaussian components) based on the animal's shell
length (SL, mm). Our algorithm also links age groups over time to form
cohorts. This unique approach is the first holistic and rigorous
analysis of age structure in Virginia's portion of Chesapeake Bay
(VACB), providing insights into the terminal age of oyster cohorts in
all the main river systems, and filling a knowledge gap in oyster
restoration approaches (\citeproc{ref-mannWhyOysterRestoration2007}{Mann
\& Powell, 2007}). Further, this analysis also suggests signals of
resiliency of VACB oyster reefs.

\phantom{xxxx}A novel GMM approach is needed for the VOSARA data for the
following reasons. Our objective is complex: we want holistic insights
at the bay-level that encapsulate river- and reef-level characteristics,
while using annual animal-level data that span two decades. Other
studies on age structure involved less complexity or employed additional
validation data that we lack. For example, Shaw et al.
(\citeproc{ref-shawCohortAnalysisEuphausia2021}{2021}) used GMM to
estimate age groups of the krill species \emph{Euphausia pacifica,} and
were able to link age groups over time to identify cohorts. However,
their study area was confined to a small region off the coast of
Newport, Oregon, with frequent sampling intervals (median of 16 days),
that allowed close monitoring of krill growth. In contrast, our VOSARA
dataset is much more spatially (222 reefs) and temporally expansive (21
years), and the sampling interval is approximately annual. Further, most
applications of GMM on size data have been used to assess how accurate
the method is compared to traditional aging methods, such as scaling of
salmonids in Sethi et al.
(\citeproc{ref-sethiAccurateAgingJuvenile2017}{2017}) , or against data
on fish with known ages and sizes
(\citeproc{ref-zhouBayesianHierarchicalApproach2020}{Zhou et al.,
2020}). No existing oyster aging method is feasible to validate the
``true'' ages of each individual oyster at the spatiotemporal scale of
VOSARA. Another major distinction between our study and previous
applications of GMM is that oysters create their own environment and are
immobile once settled
(\citeproc{ref-bonarControlOysterSettlement1990}{Bonar et al., 1990}).
Therefore, it is reasonable for our method to assume that oysters still
alive in continuing years \emph{are} part of the same cohort, because
the only way to move is through natural mortality or fishing mortality.

\phantom{xxxx}The rest of this paper is structured as follows. In
Section 2, we briefly describe the methodology of the Virginia patent
tong survey and the included data. In Section 3, we describe the
framework of our methodology to estimate the age structure of oyster
reefs in Virginia's Chesapeake Bay. This section includes the
application of GMM to live oysters and a detailed description of our
algorithm. In Section 4, we describe the algorithm output, highlighting
exemplary results from the Chesapeake Bay mainstem and James river. In
Section 5, we discuss the signals of resiliency embedded in the
estimated age structure of oysters in oyster reefs located in the
Chesapeake Bay Mainstem and James River. In Section 6, we revisit the
utility and novelty of our framework and implications of resiliency of
eastern oysters in VACB. In Section 7, we conclude by summarizing the
results of the method and its potential for future use in management.
Supplementary results from the remaining river strata are included in
the appendix.

\section{Data description}\label{data-description}

\phantom{xxxx}The VOSARA data used for this study are from quantitative
fishery independent survey data of 222 natural public oyster reefs in
the Virginia tributaries of Chesapeake Bay
(\citeproc{ref-baylorSurveyOysterGrounds1895}{Baylor, 1895}). This is a
collaborative effort between the Virginia Institute of Marine Science
(VIMS) and the Virginia Marine Resources Commission (VMRC). A
quantitative sampling program using a stratified random grid is used at
all selected reefs (\citeproc{ref-mannPopulationStudiesNative2009}{Mann
et al., 2009};
\citeproc{ref-southworthOysterCrassostreaVirginica2010}{Southworth et
al., 2010}). Oysters are collected from a hydraulic patent tong that
samples one square meter of bottom. The sampling scheme is described in
more detail in Southworth et al.
(\citeproc{ref-southworthOysterCrassostreaVirginica2010}{2010}). Our
analysis used SL data for spat and live oysters from the years 2003 to
2023 (\citeproc{ref-southworthOysterCrassostreaVirginica2010}{Southworth
et al., 2010}). Spat oysters are defined as sampled oysters that are
alive but do not have a cupped appearance. Live oysters are defined as
all other sampled oysters that are alive.

\phantom{xxxx}Data from the Nansemond River, Elizabeth River Western
Branch, Lafayette River, and Elizabeth River were excluded because they
are sampled every other year. The remaining sampled regions were
reorganized to match harvest management units set by the VMRC, which we
refer to as ``river strata''. To reduce uncertainty due to data gaps, we
excluded from reef-level analyses those reefs that did not have at least
300 oysters sampled per year over eight consecutive years or more (but
all reefs were included in stratum-level analyses). Under these
criteria, we conducted reef-level analyses on 64 reefs (n = 1,205,165
oysters across 29\% of VACB natural, public, oyster reefs). These 64
reefs covered every tributary except Lynnhaven Bay
(Figure~\ref{fig-VACB}). The river strata and the 64 included reefs are
listed in Table 1.

\small

\begin{longtable}[]{@{}
  >{\centering\arraybackslash}p{(\linewidth - 6\tabcolsep) * \real{0.4041}}
  >{\centering\arraybackslash}p{(\linewidth - 6\tabcolsep) * \real{0.1986}}
  >{\centering\arraybackslash}p{(\linewidth - 6\tabcolsep) * \real{0.1096}}
  >{\centering\arraybackslash}p{(\linewidth - 6\tabcolsep) * \real{0.2740}}@{}}
\caption{List of \((R,r)\)th river-reef strata combinations and
timeframe included in our analysis, depicted in
Figure~\ref{fig-VACB}.}\tabularnewline
\toprule\noalign{}
\begin{minipage}[b]{\linewidth}\centering
River (\(R\))
\end{minipage} & \begin{minipage}[b]{\linewidth}\centering
Reef Name
\end{minipage} & \begin{minipage}[b]{\linewidth}\centering
Reef ID (\(r\))
\end{minipage} & \begin{minipage}[b]{\linewidth}\centering
Years (\(t\))
\end{minipage} \\
\midrule\noalign{}
\endfirsthead
\toprule\noalign{}
\begin{minipage}[b]{\linewidth}\centering
River (\(R\))
\end{minipage} & \begin{minipage}[b]{\linewidth}\centering
Reef Name
\end{minipage} & \begin{minipage}[b]{\linewidth}\centering
Reef ID (\(r\))
\end{minipage} & \begin{minipage}[b]{\linewidth}\centering
Years (\(t\))
\end{minipage} \\
\midrule\noalign{}
\endhead
\bottomrule\noalign{}
\endlastfoot
\textbf{Chesapeake Bay Mainstem} & DEEP ROCK & 357 & 2003-2023 \\
& BLACKBERRY HANG & 312 & 2006-2009, 2011-2023 \\
\textbf{Great Wicomico River} & HAYNIE POINT & 64 & 2003, 2006-2023 \\
& SANDY POINT & 133 & 2003-2023 \\
& SHELL BAR & 135 & 2003-2004, 2006-2023 \\
& ROGUE POINT & 187 & 2003-2023 \\
& HARCUM FLATS & 409 & 2003-2004, 2006-2023 \\
& COE 11 & 1311 & 2006-2023 \\
& COE 13 & 1313 & 2005-2009, 2011-2023 \\
& COE 16 & 1316 & 2005-2023 \\
& HILLY WASH & 407 & 2007-2023 \\
& INGRAM'S Bay North & 385 & 2008-2009, 2011-2023 \\
\textbf{James River} & DAYS POINT & 40 & 2003-2023 \\
& POINT OF SHOALS & 123 & 2003-2023 \\
& WRECK SHOAL & 174 & 2003-2023 \\
& CROSS ROCK & 198 & 2003-2023 \\
& MULBERRY POINT & 226 & 2003-2023 \\
& UPPER DEEP WATER SHOAL & 326 & 2003-2023 \\
& LOWER DEEP WATER SHOAL & 327 & 2004, 2006-2023 \\
& UPPER HORSEHEAD & 328 & 2003-2023 \\
& MIDDLE HORSEHEAD & 329 & 2003-2023 \\
& LOWER HORSEHEAD & 330 & 2003-2023 \\
& MOON ROCK & 331 & 2003-2023 \\
& V-ROCK & 332 & 2003-2023 \\
& SHANTY ROCK & 335 & 2003-2023 \\
& DRY SHOAL & 336 & 2003-2023 \\
& SWASH & 338 & 2003-2023 \\
& UPPER JAIL ISLAND & 339 & 2003-2023 \\
& SWASH MUD SLOUGH & 340 & 2003-2023 \\
& OFFSHORE SWASH & 341 & 2003-2023 \\
& LOWER JAIL ISLAND & 342 & 2003-2008, 2010, 2012, 2014-2023 \\
& OFFSHORE JAIL ISLAND & 343 & 2003-2023 \\
& HOTEL ROCK & 346 & 2003-2005, 2007-2023 \\
& SNYDER'S ROCK & 347 & 2004-2005, 2007-2023 \\
& TRIANGLE ROCK & 348 & 2003-2023 \\
& BALLARD'S MARSH & 370 & 2006-2023 \\
& UPPER THOMAS ROCK & 373 & 2007-2008, 2010-2023 \\
& LOWER THOMAS ROCK & 374 & 2006-2023 \\
& UPPER DOG SHOAL & 375 & 2006-2008, 2010-2023 \\
& HIGH SHOAL & 377 & 2006-2023 \\
& WHITE SHOAL & 169 & 2015-2023 \\
& UPPER BROWN SHOAL & 410 & 2015-2023 \\
& LOWER BROWN SHOAL & 411 & 2015-2023 \\
\textbf{Piankatank River} & BLAND POINT & 8 & 2003, 2006-2023 \\
& BURTON POINT A & 20 & 2003, 2006-2023 \\
& CAPE TOON & 24 & 2003, 2006-2023 \\
& GINNEY POINT & 50 & 2003, 2006-2023 \\
& PALACE BAR A & 110 & 2003-2004, 2006-2023 \\
& STOVE POINT & 140 & 2006-2023 \\
& FISHING POINT & 395 & 2010-2023 \\
\textbf{Rappahannock River (includes the Corrotoman River)} & LOWER EDGE
EAST & 1004 & 2003-2023 \\
& BROAD CREEK & 1007 & 2003-2023 \\
& NORTH END & 1010 & 2003, 2006-2008, 2010-2013, 2015-2023 \\
& DRUMMING GROUND OFFSHORE & 1018 & 2005-2023 \\
& DRUMMING GROUND INSHORE & 1019 & 2003-2004, 2006-2023 \\
& BROAD CREEK INSHORE & 1028 & 2003, 2005-2023 \\
& BUTLER'S HOLE WEST & 1053 & 2008-2023 \\
& BROAD CREEK SANCTUARY & 1061 & 2007-2023 \\
& BUTLER'S HOLE SANCTUARY & 1064 & 2007-2023 \\
& BUTLER'S HOLE GRAVEL PLANT & 1067 & 2015-2023 \\
\textbf{Tangier and Pocomoke Sounds} & JOHNSON'S ROCK & 1131 &
2006-2023 \\
\textbf{York River and Mobjack Bay (includes the East River)} & ABERDEEN
ROCK & 1 & 2009-2023 \\
& PAGE'S ROCK & 109 & 2009, 2011-2023 \\
& TIMBERNECK & 389 & 2009-2023 \\
\end{longtable}

\normalsize

\begin{figure}[H]

\begin{minipage}{0.50\linewidth}

\centering{

\pandocbounded{\includegraphics[keepaspectratio]{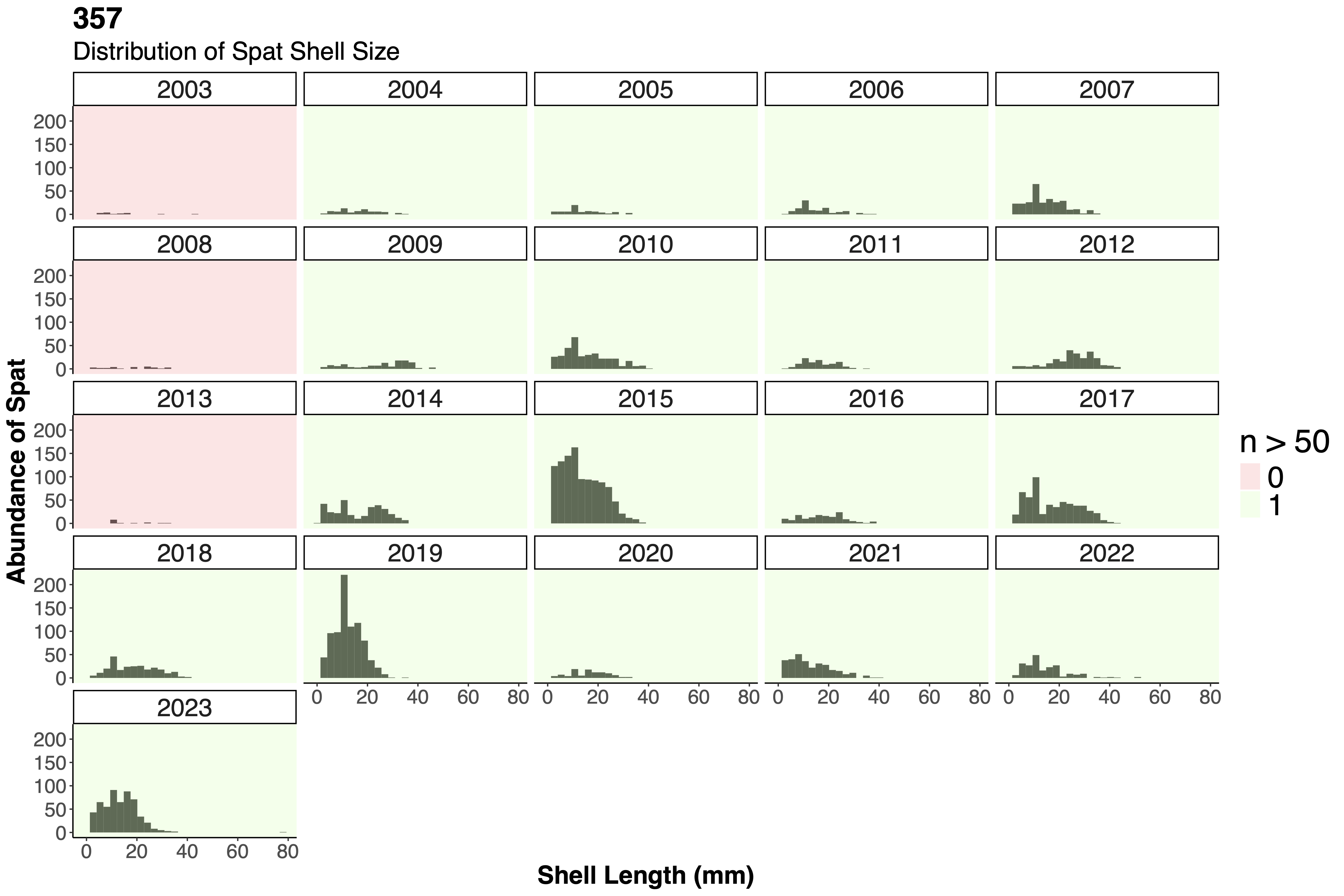}}

}

\subcaption{\label{fig-357spat}}

\end{minipage}%
\begin{minipage}{0.50\linewidth}

\centering{

\pandocbounded{\includegraphics[keepaspectratio]{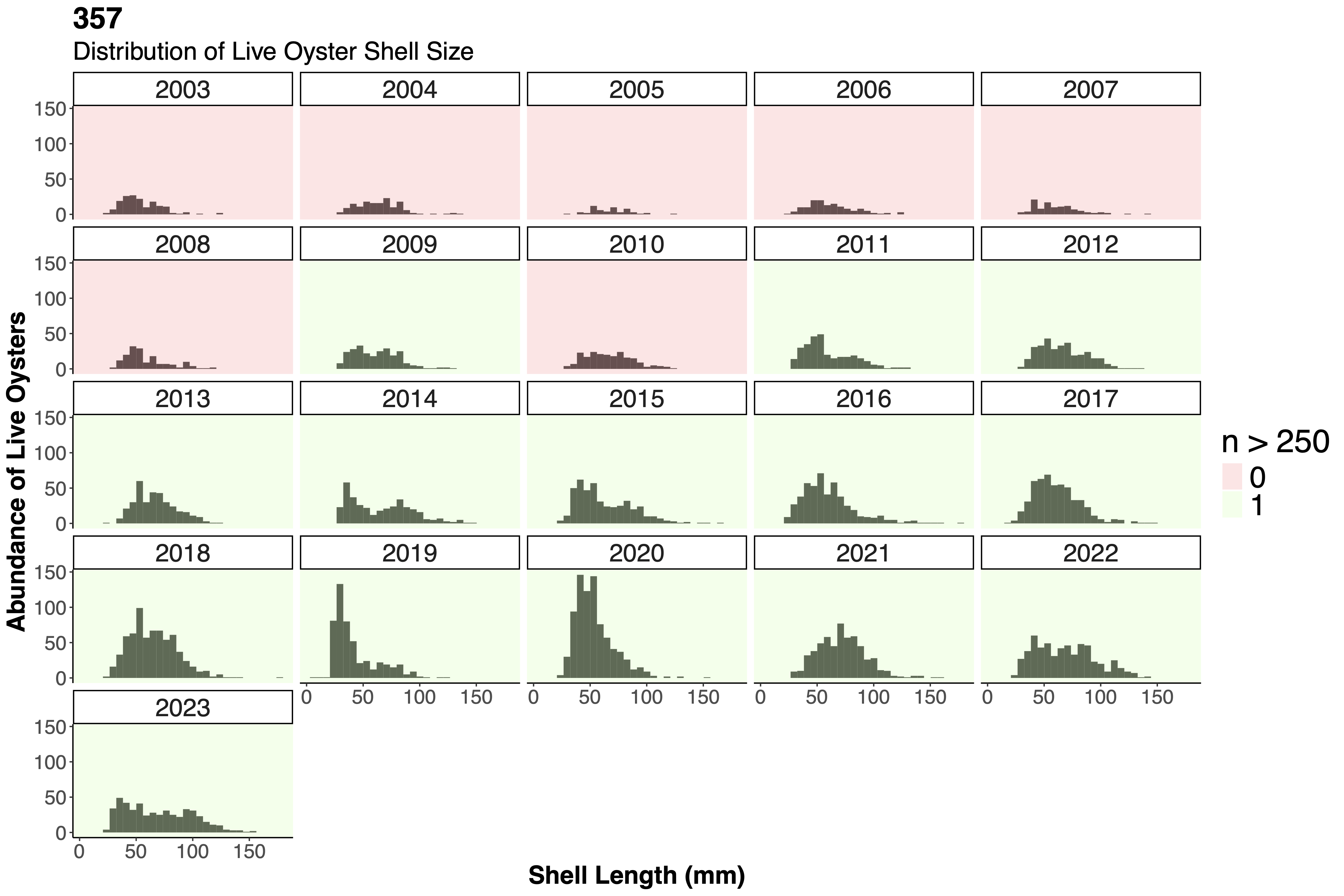}}

}

\subcaption{\label{fig-357lives}}

\end{minipage}%

\caption{\label{fig-SLdists}Annual empirical distributions of shell
length in reef 357 for spat (a) and live oysters (b).}

\end{figure}

\phantom{xxxx}After selecting included reefs, there were river-reef-year
combinations, denoted by (\(R,r,t\)), that had a sample size that was
too small for the fitting software to produce model results. As such, we
created ``spat condition'' and ``live condition'' labels for each
\((R,r,t)\)th sample (\textbf{Table 2}): a log-normal distribution was
only fit to spat oysters with at least 25 animals; a GMM was only fit to
live oysters with at least 50 animals; whenever a distribution was fit,
the sample was flagged if there were less than 50 spat or less than 250
live oysters in that sample (Figure~\ref{fig-SLdists}), suggesting that
any software output for flagged samples could be unreliable.

\small

\begin{longtable}[]{@{}
  >{\centering\arraybackslash}p{(\linewidth - 6\tabcolsep) * \real{0.2500}}
  >{\centering\arraybackslash}p{(\linewidth - 6\tabcolsep) * \real{0.2500}}
  >{\centering\arraybackslash}p{(\linewidth - 6\tabcolsep) * \real{0.2500}}
  >{\centering\arraybackslash}p{(\linewidth - 6\tabcolsep) * \real{0.2500}}@{}}
\caption{Protocol for combinations of spat and live sample sizes for
each (\(R,r,t\))th sample. *Note that the numerical computations in
\texttt{Mclust}, \texttt{MASS}, and other model-fitting software tend to
break down with too few observations.}\tabularnewline
\toprule\noalign{}
\begin{minipage}[b]{\linewidth}\centering
\end{minipage} & \begin{minipage}[b]{\linewidth}\centering
\(n_{Rrt}^{(spat)} > 50\)
\end{minipage} & \begin{minipage}[b]{\linewidth}\centering
\(25 < n_{Rrt}^{(spat)} \leq 50\)
\end{minipage} & \begin{minipage}[b]{\linewidth}\centering
\(n_{Rrt}^{(spat)} < 25\)*
\end{minipage} \\
\midrule\noalign{}
\endfirsthead
\toprule\noalign{}
\begin{minipage}[b]{\linewidth}\centering
\end{minipage} & \begin{minipage}[b]{\linewidth}\centering
\(n_{Rrt}^{(spat)} > 50\)
\end{minipage} & \begin{minipage}[b]{\linewidth}\centering
\(25 < n_{Rrt}^{(spat)} \leq 50\)
\end{minipage} & \begin{minipage}[b]{\linewidth}\centering
\(n_{Rrt}^{(spat)} < 25\)*
\end{minipage} \\
\midrule\noalign{}
\endhead
\bottomrule\noalign{}
\endlastfoot
\(n_{Rrt}^{(live)} > 250\) & \begin{minipage}[t]{\linewidth}\centering
\begin{itemize}
\item
  Log-normal distribution fit to spat SL data
\item
  GMM fit to live SL data
\end{itemize}
\end{minipage} & \begin{minipage}[t]{\linewidth}\centering
\begin{itemize}
\item
  Log-normal distribution fit to spat SL data, but flagged as a small
  sample size in data visualizations
\item
  GMM fit to live SL data
\end{itemize}
\end{minipage} & \begin{minipage}[t]{\linewidth}\centering
\begin{itemize}
\item
  No log-normal (spat data pre-emptively excluded to prevent software
  breakdown)
\item
  GMM fit to live SL data
\end{itemize}
\end{minipage} \\
\(50 < n_{Rrt}^{(live)} \leq 250\) &
\begin{minipage}[t]{\linewidth}\centering
\begin{itemize}
\item
  Log-normal distribution fit to spat SL data
\item
  GMM fit to live SL data, but flagged as a small sample size in data
  visualizations
\end{itemize}
\end{minipage} & \begin{minipage}[t]{\linewidth}\centering
\begin{itemize}
\item
  Log-normal distribution fit to spat SL data, but flagged as a small
  sample size in data visualizations
\item
  GMM fit to live SL data, but flagged as a small sample size in data
  visualizations
\end{itemize}
\end{minipage} & \begin{minipage}[t]{\linewidth}\centering
\begin{itemize}
\item
  No log-normal (spat data pre-emptively excluded to prevent software
  breakdown)
\item
  GMM fit to live SL data, but flagged as a small sample size in data
  visualizations
\end{itemize}
\end{minipage} \\
\(n_{Rrt}^{(live)} < 50\)* & \begin{minipage}[t]{\linewidth}\centering
\begin{itemize}
\item
  Log-normal distribution fit to spat SL data
\item
  No GMM, too small to reliably estimate any clusters
\end{itemize}
\end{minipage} & \begin{minipage}[t]{\linewidth}\centering
\begin{itemize}
\item
  Log-normal distribution fit to spat SL data, but flagged as a small
  sample size in data visualizations
\item
  No GMM, too small to reliably estimate any clusters
\end{itemize}
\end{minipage} & \begin{minipage}[t]{\linewidth}\centering
\begin{itemize}
\item
  No log-normal (spat data pre-emptively excluded to prevent software
  breakdown)
\item
  No GMM, too small to reliably estimate any clusters
\end{itemize}
\end{minipage} \\
\end{longtable}

\normalsize

\phantom{xxxx}Given the necessarily right skewed distribution for spat
SL (Figure~\ref{fig-357spat}) and often multimodal distribution for live
oyster SL (Figure~\ref{fig-357lives}), we fit a log-normal distribution
to spat SL data from each \((R,r,t)\)th sample and a GMM to live oyster
SL data from that same \((R,r,t)\)th sample.

\section{Methodology}\label{methodology}

\subsection{\texorpdfstring{\textbf{Overview}}{Overview}}\label{overview}

\phantom{xxxx}We fit a log-normal distribution to the spat SL data to
estimate the mean, \(\mu^{(0)}_{Rrt}\), and standard deviation,
\(\sigma^{(0)}_{Rrt}\) (see Section 3.2).

\phantom{xxxx}To estimate Gaussian components of live oysters using the
distribution of SL data, we fit initial GMMs based on how reef-level
features compared to stratum-level features of these data. Let the
estimated mean and standard deviation of SL for the \(k\)th Gaussian
component of live oysters be \(\hat\mu^{(k)}_{Rrt}\) and
\(\hat\sigma^{(k)}_{Rrt}\), respectively. By comparing
\(\hat\mu^{(k)}_{Rrt}\) to the 80th percentile of the (\(R,t\))th
distribution of SL (combining all reefs in the same river stratum), we
assigned an age \(\hat{a}^{(k)}_{Rrt}\) (ranging from two to five) to
this component and updated the initial estimates \(\hat\mu^{(k)}_{Rrt}\)
and \(\hat\sigma^{(k)}_{Rrt}\), if necessary (see Section 3.3). Then, we
developed an algorithm to link the spat and live components\footnote{The
  term ``component'' is used loosely in reference to spat data.} over
time \(t\) to identify cohorts (see Section 3.4).

\phantom{xxxx}Specifically, let \(Y_{Rrti}\) be the SL of the \(i\)th
observed oyster (spat or live) from the \((R,r,t)\)th sample. Denote the
unknown population proportions of spat and live age classes by

\[
\pi^{(k)}_{Rrt}\quad \text{for k}= k_{Rrt} = 0,1,2,..., G_{Rrt}
\]

where \(k\) = 0 denotes spat oysters, \(k\) \textgreater{} 0 denotes the
\(k\)th age class of live oysters, and \(G_{Rrt}\) is the unknown number
of live age classes. Then, our initial reef-level model for SL was

\[
Y_{Rrti} \sim \begin{cases}
\text{log-normal with } E(Y_{Rrti}) = \mu_{Rrt}^{(0)} \text{ and } SD(Y_{Rrti}) = \sigma_{Rrt}^{(0)} & \text{if } i\text{th oyster was spat},\\\mathcal{N}(\mu_{Rrt}^{(k)}, \sigma_{Rrt}^{(k)})  & \text{if } i\text{th oyster was live}\end{cases}
\]

and

\[
P(Y_{Rrti}^{(live)} \text{ is from the } k\text{th age class}) = \pi^{(k)}_{Rrt}, \quad 
\sum_{k = 1}^{G_{Rrt}}\pi_{Rrt}^{(k)} = 1-\pi_{Rrt}^{(0)}
\]

\phantom{xxxx}Thus, we assumed that each \(k\)th Gaussian component of
SLs represents an age class with \(a_{Rrt}^{(k)}\) years of age (unknown
except \(a_{Rrt}^{(0)} \equiv 1\) for spat). Age estimation involved
river-level GMM alongside the above reef-level GMM, allowing river-level
features of SL to deviate from stratum-specific structure on initial
reef-level GMM estimates. Below are details for the derivation of
\(G_{Rrt}\), parameter estimates
\(\{\hat\mu_{Rrt}^{(k)}, \hat\sigma_{Rrt}^{(k)}, \hat\pi_{Rrt}^{(k)}, \hat{a}_{Rrt}^{(k)} \}\),
and our cohort identification algorithm.

\subsection{Details of spat oyster
analysis}\label{details-of-spat-oyster-analysis}

\phantom{xxxx}First, we assigned \(\hat\pi_{Rrt}^{(0)}\) to equal the
proportion of spat relative to live oysters in each \((R,r,t)\)th
sample, such that

\[
\hat\pi_{Rrt}^{(0)} = \frac{n_{Rrt}^{(spat)}}{n_{Rrt}^{(spat)} + n_{Rrt}^{(live)}} \\
\]

\phantom{xxxx}Next, we used the \texttt{MASS:fitdistr} function to
determine \(\hat\mu_{Rrt}^{(0)}\) and \(\hat\sigma_{Rrt}^{(0)}\) for
spat SL. Finally, we assigned \(\hat{a}_{Rrt}^{(0)} = 1\) due to the
knowledge that spat are the youngest oysters of the year.

\subsection{Details of Gaussian Mixture Modeling to live oyster SL
data}\label{details-of-gaussian-mixture-modeling-to-live-oyster-sl-data}

\subsubsection{Model parameters}\label{model-parameters}

\phantom{xxxx}A mixture distribution is a probability distribution
obtained as a convex linear combination of probability density functions
(\citeproc{ref-scruccaModelBasedClusteringClassification2023a}{Scrucca
et al., 2023}). The individual distributions for the \((R,r,t)\)th
sample are mixture components labeled \(k_{Rrt}\)\emph{,} and the
weights associated with all \(G_{Rrt}\) components are mixture
probabilities \(\tilde{\pi}_{Rrt}^{(k)}\) (\(\tilde\pi_{Rrt}^{(k)} > 0\)
and \(\sum_{k = 1}^{G_{Rrt}} \tilde\pi_{Rrt}^{(k)} =  1\)) for
\(k = k_{Rrt} = 1,2,...,G_{Rrt}\). We fit the following univariate
finite Gaussian mixture distribution

\begin{equation*}
\sum_{k = 1}^{G_{Rrt}}\tilde\pi_{Rrt}^{(k)} f_{Rrt}^{(k)}(\underset{\sim{\hphantom{Y_{Rrt}}}}{Y_{Rrt}}; \mu_{Rrt}^{(k)}, \sigma_{Rrt}^{(k)})
\end{equation*}

where \(f_{Rrt}^{(k)}(.)\) is the density of the \(k\)th Gaussian
component of the \((R,r,t)\)th sample and
\(\underset{\sim{\hphantom{Y_{Rrt}}}}{Y_{Rrt}}\) is the vector of shell
lengths of live oysters.

\phantom{xxxx}We fit this model, with no specified covariance structure,
using the \texttt{mclust} package
(\citeproc{ref-scruccaModelBasedClusteringClassification2023a}{Scrucca
et al., 2023}). We allow the \texttt{mclust::Mclust} function to decide
the value for \(G_{Rrt}\) from 1 to 4, reflecting our current
understanding that four-year-old oysters (in their 5th year) are rare in
Chesapeake Bay
(\citeproc{ref-hardingManagementPiankatankRiver2010}{Harding et al.,
2010};
\citeproc{ref-southworthOysterCrassostreaVirginica2010}{Southworth et
al., 2010}). The \texttt{Mclust} function uses an
expectation-maximization algorithm and the Bayesian information
criterion (BIC) to select the best model and best fit of \(G_{Rrt}\).
Scrucca et al.
(\citeproc{ref-scruccaModelBasedClusteringClassification2023a}{2023})
and Neath \& Cavanaugh
(\citeproc{ref-neathBayesianInformationCriterion2012}{2012}) recommend
that \(\Delta_{BIC} \gtrapprox 6\) suggests strong evidence in favor of
one model over the other. The \texttt{mclust::Mclust} function
automatically selects the best fitting model (largest BIC), regardless
of the amount of evidence in favor of the ``best model''. As such, we
manually selected the more parsimonious model when there was weak
evidence in favor of one model over the other (\(\Delta_{BIC} < 2\)).

\phantom{xxxx}We posit that each component \(k_{Rrt}\) should have
varying variances (represented by a ``V'' model), as opposed to an ``E''
model with equal variances (see documentation in Scrucca et al.
(\citeproc{ref-scruccaModelBasedClusteringClassification2023a}{2023}))
unless there is strong evidence against it. As such, for each
(\(R,r,t\))th GMM where \(\Delta_{BIC} < 2\), we extracted the BIC of
all models (\(1 \le G_{Rrt} \le 4\)) of varying variance ``V'' . We
calculated \(\Delta_{BIC}\) between all model fits with varying variance
``V'' and the ``best-fitting'' model determined by
\texttt{mclust::Mclust}. For the ``V'' models whose \(\Delta_{BIC} < 2\)
when compared to the ``best-fitting'' model, we selected the ``V'' model
with the smallest \(G_{Rrt}\). If there was no ``V'' model that met
these criteria, we repeated the process to pick the most parsimonious
``E'' model.

\phantom{xxxx}The \texttt{Mclust} function sets
\(\sum_{k = 1}^{G_{Rrt}} \hat{\tilde\pi}_{Rrt}^{(k)} = 1\). Since each
\((R,r,t)\)th sample includes spat (\(k\) = 0) and live components
(\(k = k_{Rrt} = 1,2,...,G_{Rrt}\)), we adjusted
\(\hat{\tilde\pi}^{(k)}_{Rrt}\) by taking

\[
\hat{\pi}_{Rrt}^{(k)} =  \hat{\tilde\pi}^{(k)}_{Rrt}(1 - \hat\pi_{Rrt}^{(0)})
\]

so that

\begin{align*}
\sum_{k = 1}^{G_{Rrt}} \hat\pi_{Rrt}^{(k)} = (1 - \hat\pi_{Rrt}^{(0)}) = \frac{n_{Rrt}^{(live)}}{n_{Rrt}^{(spat)} + n_{Rrt}^{(live)}}
\end{align*}

\phantom{xxxx}The adjusted mixture weights \(\hat\pi_{Rrt}^{(k)}\)'s
were then used for cohort identification in Section 3.4.

\subsubsection{Age estimation and updates to reef-level
estimates}\label{age-estimation-and-updates-to-reef-level-estimates}

\phantom{xxxx}Given that the shape and size of oysters tend to vary
based on environment (typically at a stratum-level as opposed to a
reef-level (\citeproc{ref-marquardtOysterAllometryGrowth2024}{Marquardt
et al., 2024})), each \(k_{Rrt}\)th (reef-level) Gaussian component was
estimated to have age \(\hat{a}_{Rrt}^{(k)}\) based on its comparison to
river-level data, as follows. First we combined spat and live oysters
sampled from the same \((R,t)\)th combination (fixed river stratum,
fixed year) \emph{i.e.,}
\(\underset{\sim{\hphantom{Y_{Rt}}}}{Y_{Rt}} = \{Y_{Rrti}: \text{all }i\text{ and all } r \text{ from } R \text{ even if } r \text{ was excluded from reef-level analyses}\}\).
We used \texttt{Mclust} to fit the river-level Gaussian mixture model

\begin{equation*}
\sum_{m = 1}^{G_{Rt}}\pi_{Rt}^{(m)} f_{Rt}^{(m)}(\underset{\sim{\hphantom{Y_{Rt}}}}{Y_{Rt}}; \mu_{Rt}^{(m)}, \sigma_{Rt}^{(m)})
\end{equation*}

where \(G_{Rt}\) is the number of mixture components
(\(1 \le G_{Rt} \le 4\)), \(f_{Rt}^{(m)}(.)\) is the density of the
\(m\)th Gaussian component of the \((R,t)\)th river stratum (with \(m\)
= 1, \ldots, \(G_{Rt}\)), the \(\pi_{Rt}^{(m)}\)'s are the mixture
weights (\(\pi_{Rt}^{(m)} > 0\) and
\(\sum_{m = 1}^{G_{Rt}} \pi_{Rt}^{(m)} =  1\)), and \(\mu_{Rt}^{(m)}\)
and \(\sigma_{Rt}^{(m)}\) are the Gaussian parameters of the \(m\)th
density component. Similarly to the GMMs at the (\(R,r,t\))th level, we
manually selected the more parsimonious model when there was weak
evidence of difference between models (\(\Delta_{BIC} < 2\)).

\phantom{xxxx}This Gaussian mixture model at the \((R,t)\)th level
allows us to employ our understanding of the age structure at each
\(R\)th river stratum. We determined that the 80th quantile was an
appropriate cutoff for estimating ages\footnote{Preliminary sensitivity
  analyses (not shown) suggested robustness of our method to the choice
  of this cutoff.}, the youngest age set to 2 (\emph{i.e.,} a year older
than spat). The estimation mechanism is as follows:

\begin{align*} 
&\bullet \quad \text{} && q_{Rt}^{(0)} = 0 \\
&\bullet \text{ Fix }R,r, \text{ and } m. \text{ Determine} && q_{Rt}^{(m)} = \text{80th percentile of } \mathcal{N}(\hat\mu_{Rt}^{(m)}, \hat\sigma_{Rt}^{(m)}) \text{ for } m > 0 \\
&\bullet \text{ For each } k_{Rrt}=1,...,G_{Rrt}\text{: if} && q_{Rt}^{(m-1)} < \hat\mu_{Rrt}^{(k)} \le q_{Rt}^{(m)} \text{ then} \quad \hat{a}_{Rrt}^{(k)} := m + 1 \\
&\bullet \text{ Repeat over all }(R,t,m).
\end{align*}

\phantom{xxxx}When \(G_{Rt} = 1\), \(\hat{a}_{Rrt}^{(k)}\) is assigned
``N.A.'', meaning that there was no useable knowledge about the size of
live age groups (age 2+) in that river system, and as such, we could not
reasonably assign \(\hat{a}_{Rrt}^{(k)}\).

\phantom{xxxx}Next, initial reef-level \(\hat\mu_{Rrt}^{(k)}\) and
\(\hat\sigma_{Rrt}^{(k)}\) from Section 3.3.1 were subject to updating
based on GMM fits to reef-level data, whenever
\(\hat{a}_{Rrt}^{(k)} = \hat{a}_{Rrt}^{(k')}\) for \(k \neq k'\). The
reason is as follows.

\phantom{xxxx}In an \((R,r,t)\)th population, the true age
\(a_{Rrt}^{(k)}\) cannot be identical for different values of \(k\) if
age dictates SL (\emph{i.e.,} at the same reef in the same year, two
separate Gaussian components cannot have the same age). Although
recruitment could have continued after each sampling date, by the time
sampling occurs the following fall, all spat oysters from the previous
sample tend to be aggregated into similar sizes identifiable as live
oysters. Therefore, we ignored possible new recruits among live oysters
and calculated the following updated estimates to pool those components
sharing the same age assignment into one:

\begin{gather*}
\hat\mu_{pooled} = \sum_{k = 1}^{p_{Rrt}}\hat{\tilde\pi}^{(k)}_{Rrt}*\hat\mu^{(k)}_{Rrt} \\\\
\hat\sigma^{2}_{pooled} = \left[\sum_{k = 1}^{p_{Rrt}}\hat{\tilde\pi}^{(k)}_{Rrt}(\hat\sigma^{2(k)}_{Rrt}+\hat\mu^{2(k)}_{Rrt})\right] - \hat\mu^2_{pooled}
\end{gather*}

where \(p_{Rrt}\) is the number of components in the same \((R,r,t)\)th
sample showing the same value for \(\hat{a}_{Rrt}^{(k)}\). In these
instances, \(\hat\mu_{pooled}\) and \(\hat\sigma_{pooled}\) were used in
the cohort assignment algorithm.

\subsection{Cohort assignment
algorithm}\label{cohort-assignment-algorithm}

\phantom{xxxx}The following algorithm was used to link spat and live
components over time to form age cohorts (\(\hat{c}^{(k)}_{Rrt}\)).

\begin{center}\rule{0.5\linewidth}{0.5pt}\end{center}

\textbf{Algorithm 1: Cohort Assignment Algorithm for determination of
cohort label} \(\hat{c}^{(k)}_{Rrt}\)\footnote{Note that our
  implementation of this algorithm employed a different set of labels.}

\small

\textbf{Input:} Dataframe \(\mathbb{D}\) with column headings
\(\{R, r,t, k, \hat\mu^{(k)}_{Rrt}, \hat{a}^{(k)}_{Rrt}, \hat{c}^{(k)}_{Rrt}\}\)
collating all component means and their estimated ages after any
necessary pooling. \(\quad\) \(\texttt{// All values of}\hspace{0.5em}\)
\(\hat{c}^{(k)}_{Rrt}\) \(\texttt{ initiated  as "N.A."}\)

\vspace{0.5em}

\begin{enumerate}
\def\labelenumi{\arabic{enumi}.}
\tightlist
\item
  \textbf{Assign} column
  \(\hat{c}^{(k)}_{Rrt} := \texttt{"}t.\hat{a}_{Rrt}^{(k)}\texttt{"}\)
  \(\qquad\)
  \(\texttt{// Unique unless}\hspace{0.5em} \hat{a}^{(k)}_{Rrt} := \texttt{"N.A."}\)
\end{enumerate}

\vspace{0.5em}

\begin{enumerate}
\def\labelenumi{\arabic{enumi}.}
\setcounter{enumi}{1}
\tightlist
\item
  \textbf{for} \(R = 1, 2, ..., 7\) \textbf{do} \(\qquad\)
  \(\texttt{// River stratum label}\)
\end{enumerate}

\vspace{0.5em}

\begin{enumerate}
\def\labelenumi{\arabic{enumi}.}
\setcounter{enumi}{2}
\tightlist
\item
  \(\quad\)\textbf{for} \(r = 1, 2, ... n_R\) \textbf{do} \(\qquad\)
  \(\texttt{// reef label,} \hspace{0.5em} n_R \hspace{0.5em} \texttt{in total}\)
\end{enumerate}

\vspace{0.5em}

\begin{enumerate}
\def\labelenumi{\arabic{enumi}.}
\setcounter{enumi}{3}
\tightlist
\item
  \(\quad\quad\)\textbf{for} \(t = 2003, 2004, ..., 2022\) \textbf{do}
\end{enumerate}

\(\quad\quad\quad\quad\quad\) \textbf{4.1. Extract} from \(\mathbb{D}\)
the (\(R,r,t\))th dataframe \(\mathbb{D}_{Rrt}\) with column headings
\(\{R, r, t,k, \hat\mu^{(k)}_{Rrt}, \hat{a}^{(k)}_{Rrt}, \hat{c}^{(k)}_{Rrt} \}\)

\(\quad\quad\quad\quad\quad\) \textbf{4.2. Extract} from \(\mathbb{D}\)
the (\(R,r,t+1\))th dataframe \(\quad\)\(\mathbb{D}_{R,r,t+1}\) with
column headings
\(\{R, r, t+1, k', \hat\mu^{(k')}_{R,r,t+1}, \hat{a}^{(k')}_{R,r,t+1}, \hat{c}^{(k')}_{Rrt} \}\)

\vspace{0.5em}

\begin{enumerate}
\def\labelenumi{\arabic{enumi}.}
\setcounter{enumi}{4}
\tightlist
\item
  \(\quad\quad\quad\)\textbf{for} each \(k\)th row in
  \(\mathbb{D}_{Rrt}\) \textbf{do}
\end{enumerate}

\(\quad\quad\quad\quad\quad\quad\) \textbf{5.1. Extract} (\(k'\))th row
from \(\mathbb{D}_{R,r,t+1}\) for which
\(\hat{a}^{(k')}_{R,r,t+1} == \hat{a}^{(k)}_{R,r,t} + 1\) \(\qquad\)
\(\texttt{// can be} \hspace{0.5em} \emptyset\)

\(\quad\quad\quad\quad\quad\quad\) \textbf{5.2. if} 5.1 is not
\(\emptyset\) \textbf{then}

\(\quad\quad\quad\quad\quad\quad\quad\) \textbf{5.2.1. if}
\(\hat\mu^{(k')}_{R,r,t+1} > \hat\mu^{(k)}_{R,r,t}\) \textbf{then
update} \(\hat{c}_{R,r,t+1}^{(k')} := \hat{c}_{R,r,t}^{(k)}\) \(\qquad\)
\(\texttt{// same cohort}\)

\vspace{0.5em}

\begin{enumerate}
\def\labelenumi{\arabic{enumi}.}
\setcounter{enumi}{5}
\tightlist
\item
  \(\quad\quad\) \textbf{end} loop \(t\) \(\qquad\)
  \(\texttt{//} \quad \mathbb{D} \hspace{0.5em} \texttt{is now updated with new} \hspace{0.5em} \mathbb{D}_{R,r,t+1}\)
\end{enumerate}

\vspace{0.5em}

\begin{enumerate}
\def\labelenumi{\arabic{enumi}.}
\setcounter{enumi}{6}
\tightlist
\item
  \(\quad\) \textbf{end} loop \(r\)
\end{enumerate}

\vspace{0.5em}

\begin{enumerate}
\def\labelenumi{\arabic{enumi}.}
\setcounter{enumi}{7}
\tightlist
\item
  \textbf{end} loop \(R\)
\end{enumerate}

\normalsize

\begin{center}\rule{0.5\linewidth}{0.5pt}\end{center}

\phantom{xxxx}After running the algorithm, any components labeled
\(\texttt{"}N.A.\texttt{"}\) in column \(\hat{c}^{(k)}_{Rrt}\), were
given unique labels (\(\texttt{"N.A.1"}\)\emph{,} \(\texttt{"N.A.2"}\)
\emph{,} \ldots{} \emph{,} \(n_{N.A.}\)) where \(n_{N.A.}\) is the
number of components assigned \(\texttt{"}N.A.\texttt{"}\) in each
(\(R,r,t\))th sample.

\section{Algorithm output}\label{algorithm-output}

\begin{figure}[H]

\centering{

\pandocbounded{\includegraphics[keepaspectratio]{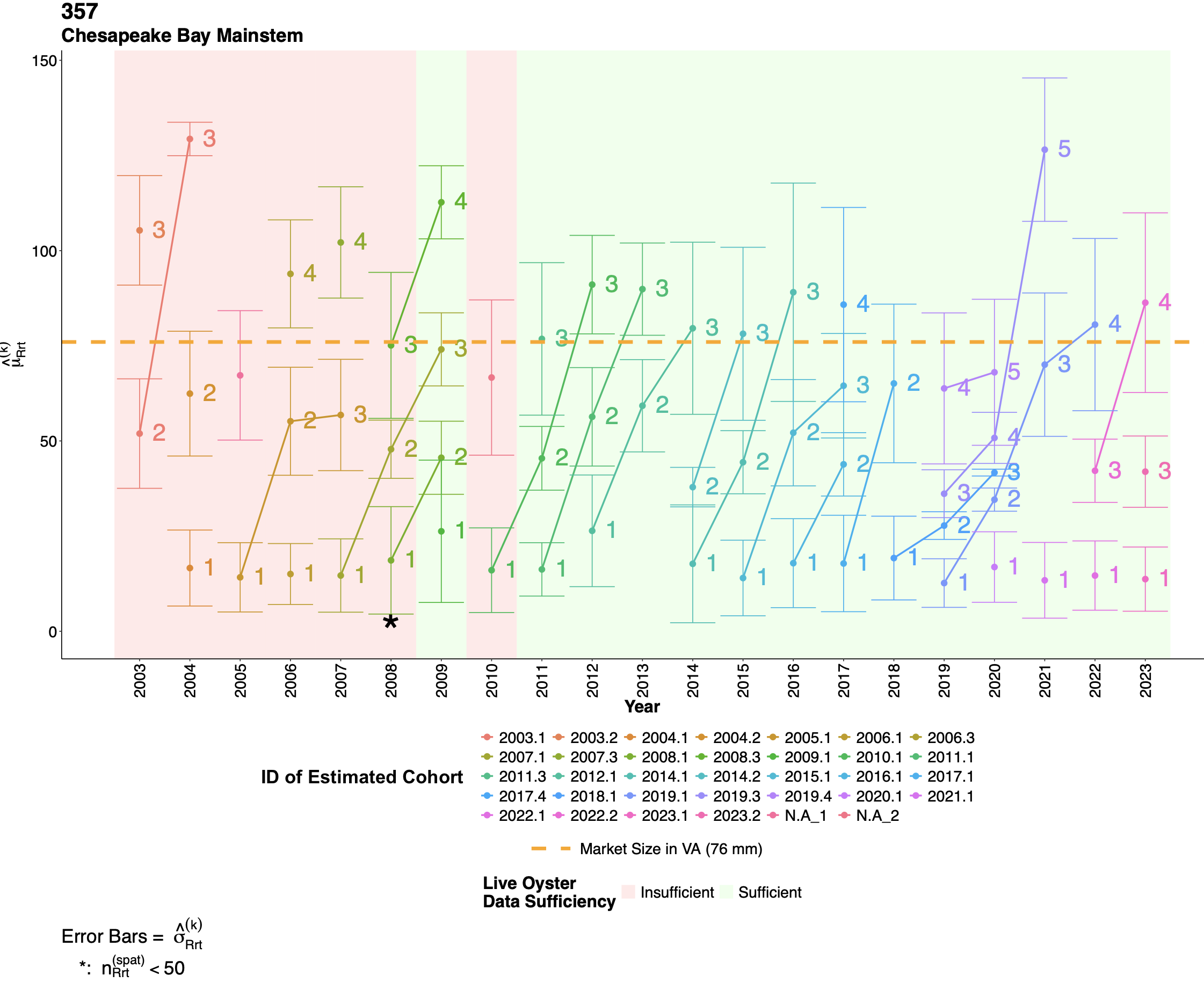}}

}

\caption{\label{fig-357}Temporal visualization of estimated cohorts
\(\hat{c}^{(k)}_{Rrt}\) in Reef 357 in the Chesapeake Bay Mainstem.
Points represent \(\hat{\mu}^{(k)}_{Rrt}\) and error bars represent
\(\hat\sigma^{(k)}_{Rrt}\) of the estimated components. A green panel
background represents \(n_{Rrt}^{(live)} > 250\), while a pink panel
background represents \(n_{Rrt}^{(live)} \leq 250\). A black asterisk
(\(*\)) represents \(n_{Rrt}^{(spat)} < 50\). The number next to each
point represents \(\hat{a}^{(k)}_{Rrt}\). The orange dotted line
represents the market size in Virginia (76 mm). The figures for all the
reefs analyzed can be found in Appendix X.}

\end{figure}

\begin{figure}[H]

\centering{

\pandocbounded{\includegraphics[keepaspectratio]{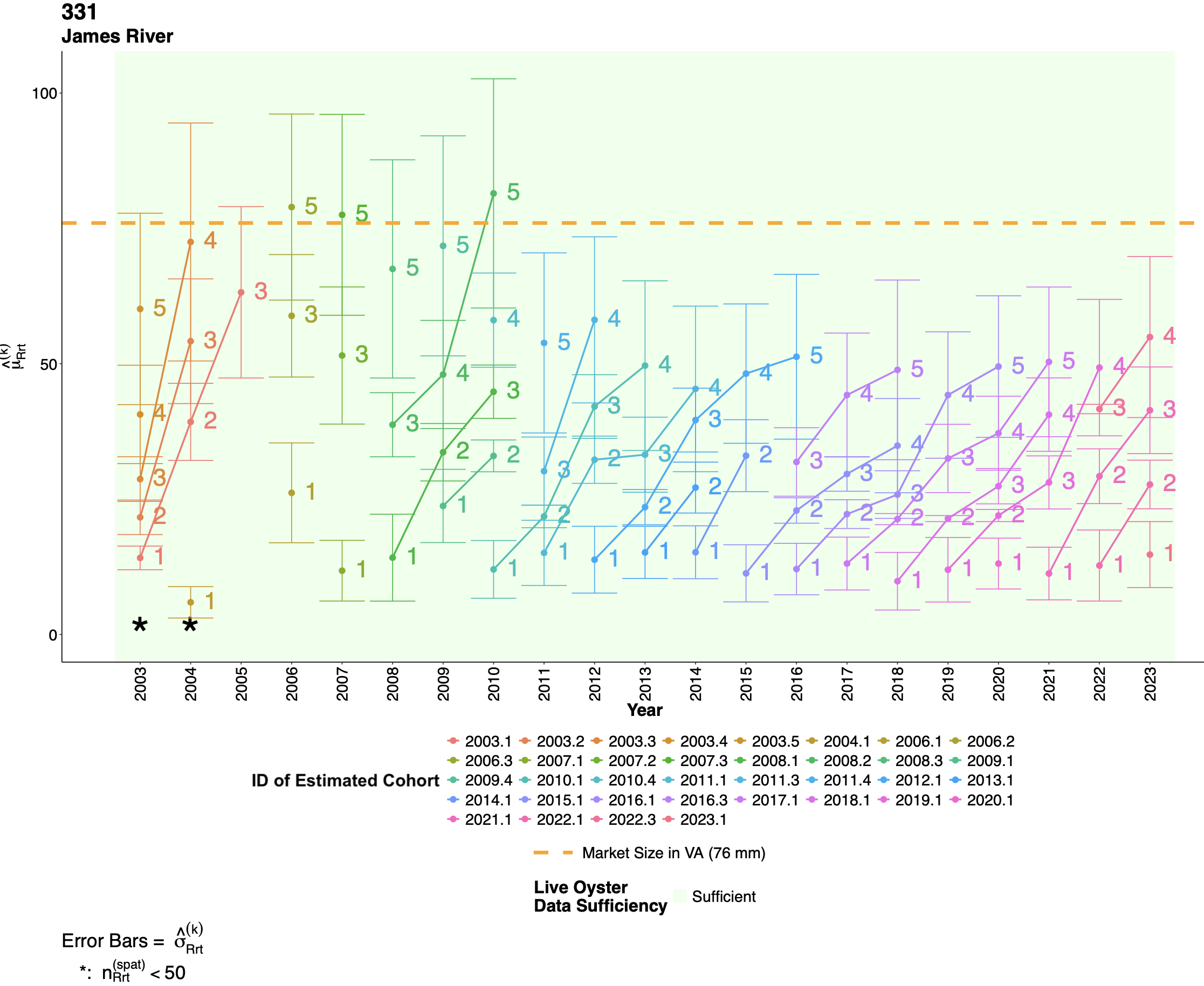}}

}

\caption{\label{fig-331}Temporal visualization of each algorithm output
for Reef 331 in the James River.}

\end{figure}

\phantom{xxxx}First, we discuss the algorithm output for two reefs,
namely reef 357 in the Chesapeake Bay Mainstem (Figure~\ref{fig-357})
and reef 331 in the James River (Figure~\ref{fig-331}). The discussion
illustrates the utility of our approach.

\phantom{xxxx}For the first few years of sampling (\(t\) = 2003-2008,
2010), reef 357 had an insufficient amount of data (pink years and/or
asterisks in Figure~\ref{fig-357}) and therefore a small number of age
groups, many of which could not be linked. During this time frame,
recruitment (shown by age one, \emph{i.e.,} spat groups) was consistent,
suggesting that spat formed the dominant age group during those years.
As time progressed, live oysters were more abundant on the reef, spat
(and therefore recruitment) continued to be consistent, and cohort
lifespan began to lengthen, shown by the presence of older age groups.
This same trend was found in reef 312, the other reef analyzed in the
Chesapeake Bay Mainstem, though reef 312 had more years with
insufficient amounts of data. In reef 357, there were two cohorts whose
terminal age was five, compared to one in reef 312, as of 2023. Most
oysters were smaller than the market size in Virginia (generally set at
76 mm throughout this time period), shown by the few error bars above
the orange dotted line. In the latter decade, cohorts could be tracked
for longer periods and were living longer (Figure~\ref{fig-357}). In the
last few years, oysters may or may not grow to four or five years old,
because we do not have the data after 2023 to estimate their lifespan.
The mean shell size of spat oysters were relatively stable across the
entire time period.

\phantom{xxxx}Reef 331 in the James River (Figure~\ref{fig-331}) was
sampled consistently throughout the entire time frame, though from 2003
to 2005, recruitment was poor, as suggested by the lack of reliable
log-normal parameter estimates due to insufficient spat data. Older age
cohorts were found throughout the time series as well. Many of the age
groups could be linked together, suggesting longevity of cohorts. All
error bars after 2010 were below market size in Virginia. This pattern
is common in all reefs (\(n_R = 31\)) in the James River (Appendix).

\phantom{xxxx}The patterns shown in reef 357 and reef 331 support the
findings of differences in size and growth of oysters in different river
systems by Marquardt et al.
(\citeproc{ref-marquardtOysterAllometryGrowth2024}{2024}). In general,
oysters in the James River are smaller than oysters in the Chesapeake
Bay Mainstem, but both exhibit signals of reduced age-truncation despite
signals of shell shrinkage. Oysters in the James River more consistently
lived to an estimated four and five years old compared to other
tributaries in Virginia's Chesapeake Bay. The lines connecting
components into estimated cohorts tend to be longer and less steep for
reef 331 then for reef 357, indicating longer-lived, but slower-growing
cohorts for this and other reefs in the James (given the similar
patterns across the James reefs we analyzed).

\begin{figure}[H]

\centering{

\pandocbounded{\includegraphics[keepaspectratio]{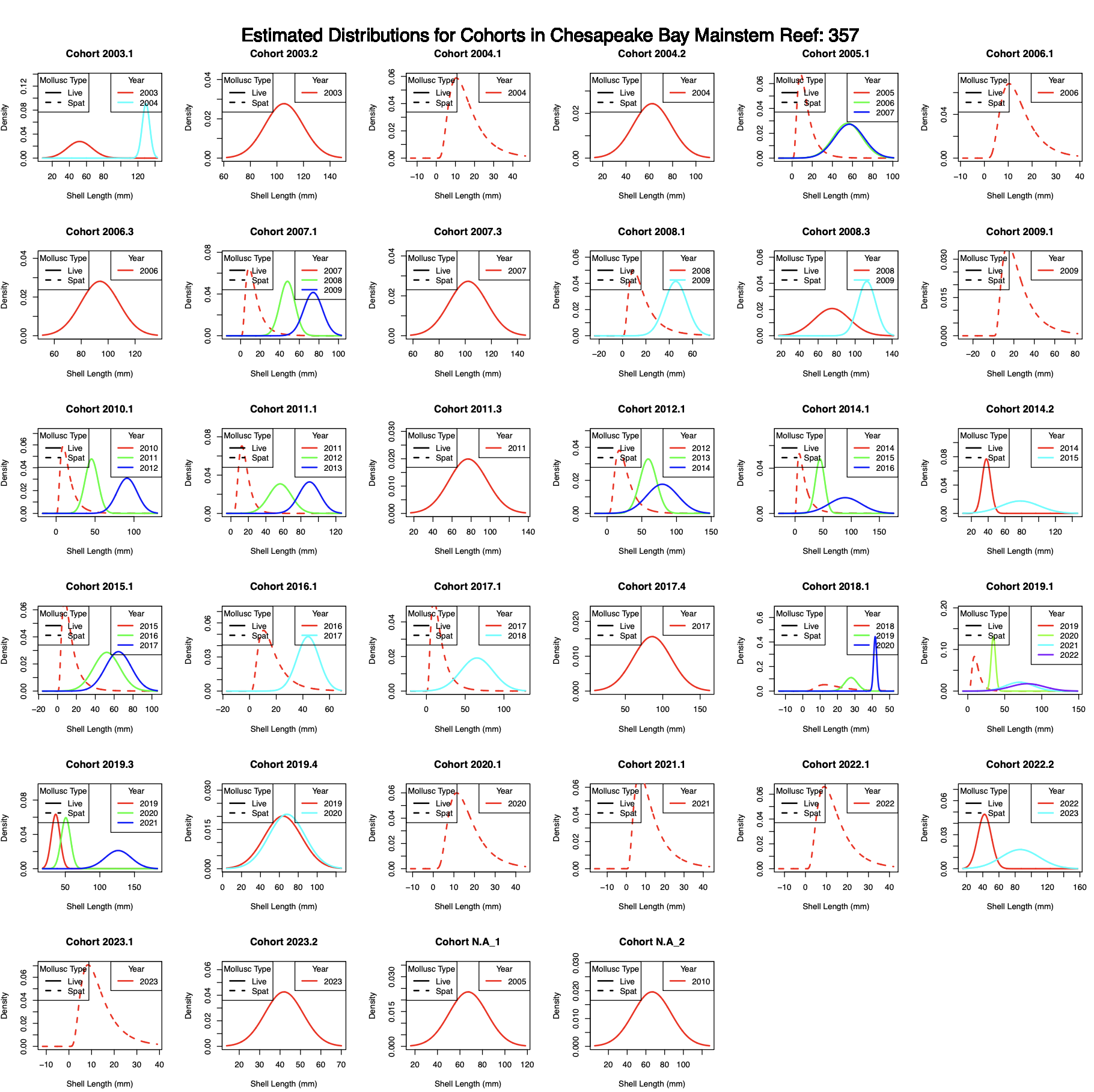}}

}

\caption{\label{fig-overlap}Estimated distributions for cohorts in
Chesapeake Bay Mainstem reef 357. Each pane refers to a different
identified cohort (\(\hat{c}_{Rrt}^{(k)}\)). Note that the mixture
weights of each cohort (\(\hat\pi_{Rrt}^{(k)}\)) do not sum to one
because each component \(k\) is linked over time, while the GMM (with
its mixture weights) was estimated yearly.}

\end{figure}

\phantom{xxxx}The above results for these reefs suggest that cohorts
could be reasonably estimated using our method. When the year-specific
distributions for each cohort in reef 357 (Chesapeake Bay Mainstem) are
combined in a single display (as a pane in Figure~\ref{fig-overlap}),
distributions rarely overlap much, allowing a clear temporal tracking of
growth in size, while relative abundance \(\hat\pi^{(k)}_{Rrt}\)
decreased, possibly due to oysters dying off as they aged. The
distributions of all cohorts identified for all 64 reefs can be found in
the Appendix.

\section{Signals of resiliency}\label{signals-of-resiliency}

\subsection{Age-truncation}\label{age-truncation}

\begin{figure}[H]

\begin{minipage}{0.50\linewidth}

\centering{

\pandocbounded{\includegraphics[keepaspectratio]{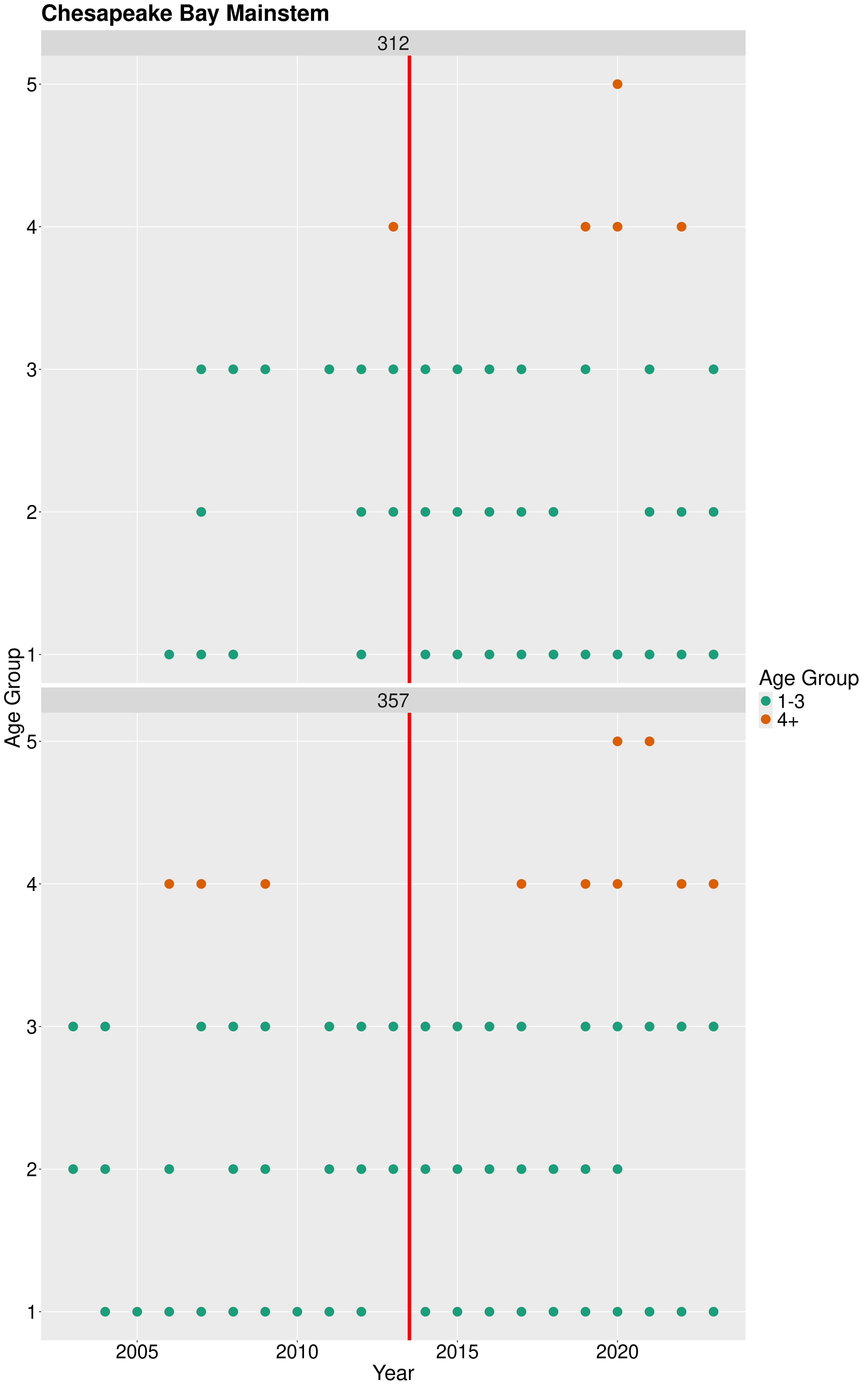}}

}

\subcaption{\label{fig-CBdot}Chesapeake Bay Mainstem}

\end{minipage}%
\begin{minipage}{0.50\linewidth}

\centering{

\pandocbounded{\includegraphics[keepaspectratio]{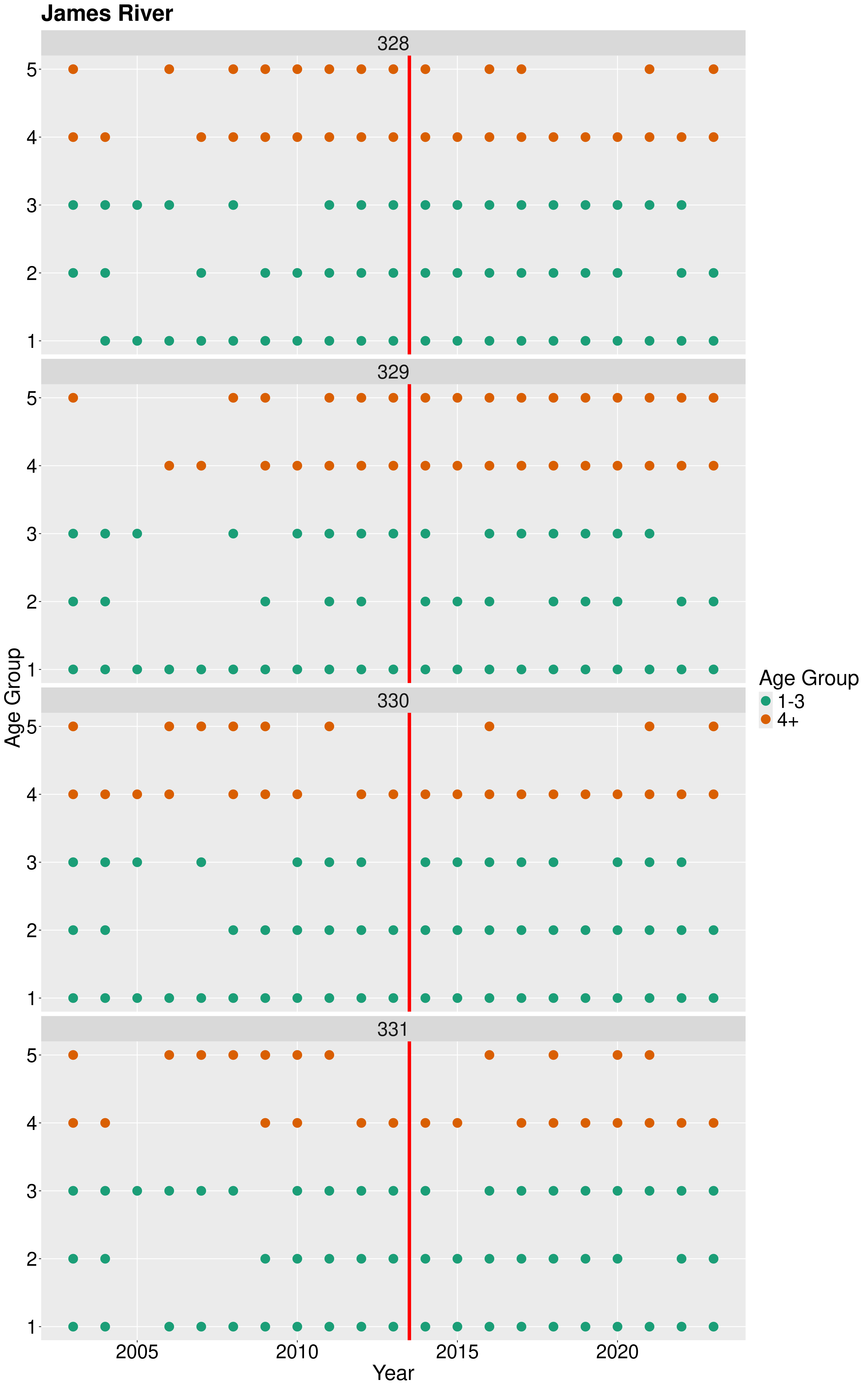}}

}

\subcaption{\label{fig-Jamesdot}James River}

\end{minipage}%

\caption{\label{fig-dots}Age structure over time in each analyzed reef
from the Chesapeake Bay Mainstem (a) and select analyzed reefs from the
James River (b). The red line represents the midpoint in the 21-year
time series (2013). Green dots represent estimated ages of one to three
and orange dots four and older.}

\end{figure}

\phantom{xxxx}For almost all reefs in our analysis, older age groups
(\emph{i.e.,} \(\hat{a}_{Rrt}^{(k)} \ge 4\)) were more present in the
second decade of the time series, suggesting that age-truncation issues
was improving in Virginia's Chesapeake Bay. In the Chesapeake Bay
Mainstem (Figure~\ref{fig-CBdot}), older age groups were present more
frequently in the latter decade (2014-2023) compared to the first
decade. For reef 312, a four-year-old age group was not estimated until
2019, then found almost every year after. Further, in any latter year,
the age structure was more diverse, meaning that there were more age
groups living on a reef at the same time. This trend was seen in all of
Virginia's tributaries, except in the James River (see Appendix). In the
James River (Figure~\ref{fig-Jamesdot}), older age groups were
consistently present across the time series, and a multi-year-class age
structure was much more common.

\subsection{Shell size shrinkage}\label{shell-size-shrinkage}

\begin{figure}[H]

\begin{minipage}{0.50\linewidth}

\centering{

\pandocbounded{\includegraphics[keepaspectratio]{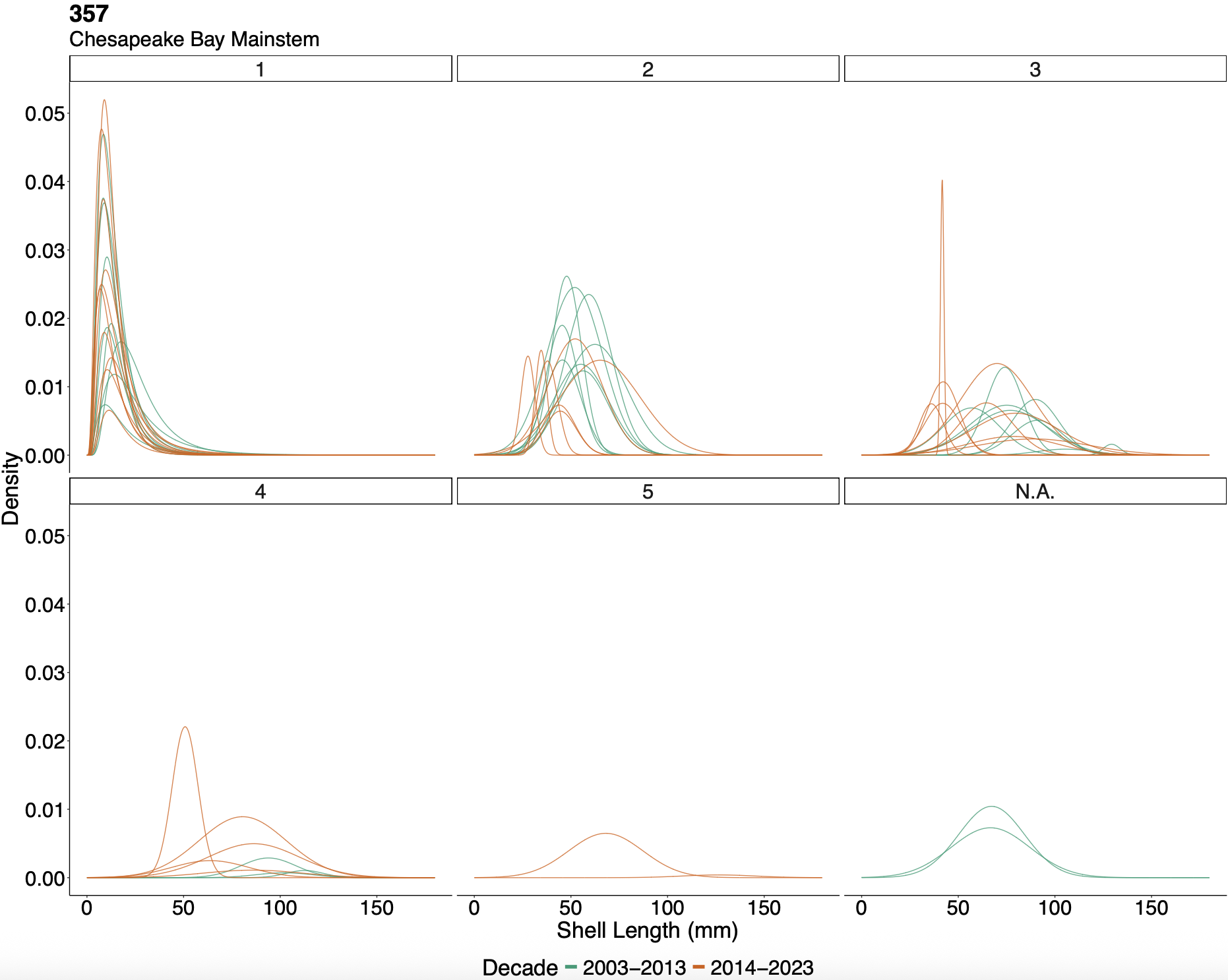}}

}

\subcaption{\label{fig-CBshrink}Chesapeake Bay Mainstem}

\end{minipage}%
\begin{minipage}{0.50\linewidth}

\centering{

\pandocbounded{\includegraphics[keepaspectratio]{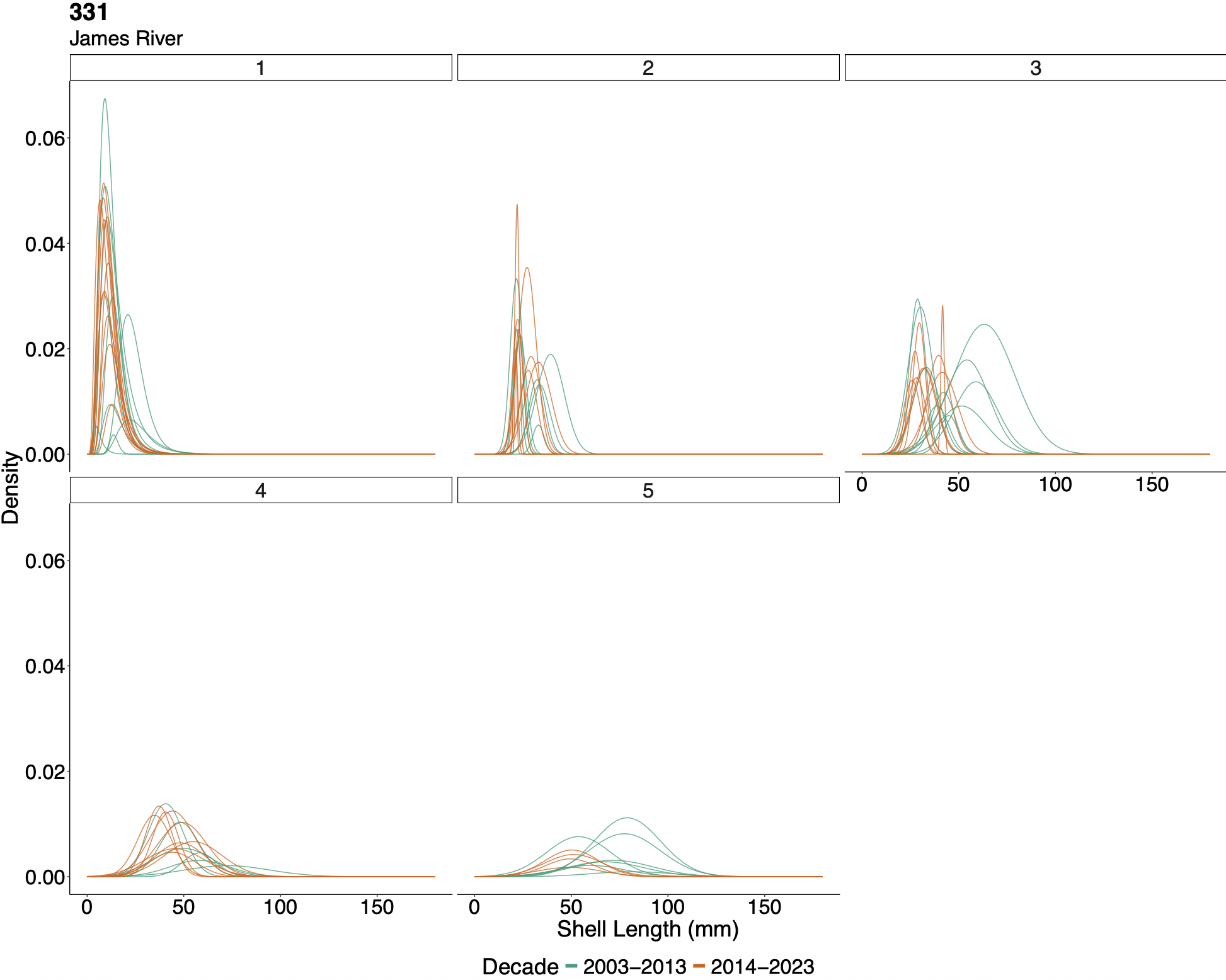}}

}

\subcaption{\label{fig-Jamesshrink}James River}

\end{minipage}%

\caption{\label{fig-shrink}SL distributions of each estimated age group,
colored by decade, for Chesapeake Bay Mainstem reefs (a) and selected
reefs in the James River (b). The facet labels represent each
\(\hat{a}_{Rrt}^{(k)}\). Figures for all 64 reefs can be found in the
Appendix.}

\end{figure}

\phantom{xxxx}When comparing the distribution of SL of estimated age
groups (\(\hat{a}^{(k)}_{Rrt}\)) between decades, we found evidence of
shell size shrinkage over time. Within the same estimated age group,
estimated distributions of SL were made up of larger values in the first
decade compared to the second, shown in green and brown, respectively,
in Figure~\ref{fig-shrink}. This trend was found in all analyzed reefs,
but strongest in Chesapeake Bay Mainstem (Figure~\ref{fig-CBshrink}) and
James river reefs (Figure~\ref{fig-Jamesshrink}). For both river
systems, this trend was less apparent for spat (\emph{i.e.},
\(\hat{a}^{(k)}_{Rrt} = 1\)).

\section{Discussion}\label{discussion}

\subsection{Method utility}\label{method-utility}

\phantom{xxxx}This research demonstrates a practical method for
identifying age groups and linking them into cohorts solely based on SL
data observed across space and time. This method allows one to conduct
an age-structure analysis at the most spatially granular level (oyster
reefs, \(r\)), while incorporating information about the corresponding
river system or tributary (\(R\)), facilitating the comparison of the
age structure and oyster shell size across river systems. Though
consistent sampling for consecutive years (here, we considered at least
eight consecutive years) is required, this method could be applied to
other managed species that do not have adequate age data, and
age-structured management approaches are desired. Future studies could
apply this method to analyze growth, mortality, and shell accretion
rates similar to previous studies in Chesapeake Bay
(\citeproc{ref-hardingManagementPiankatankRiver2010}{Harding et al.,
2010}; \citeproc{ref-mannPopulationStudiesNative2009}{Mann et al.,
2009}; \citeproc{ref-marquardtOysterAllometryGrowth2024}{Marquardt et
al., 2024};
\citeproc{ref-southworthOysterCrassostreaVirginica2010}{Southworth et
al., 2010}). Though practical for estimating and identifying age groups
based on SL data only, the method has limitations. In a few instances, a
live age group (\(\hat{a} \ge 2\)) would not be linked to any spat group
(\(\hat{a} = 1\)), which is biologically illogical.

\phantom{xxxx}For our complex objective of gaining bay-level insights
informed by river- and reef-level understanding---all based on
animal-level data---a fully rigorous methodology may be a Bayesian
hierarchical model (BHM) that captures bay-wide features through
bay-level model parameters, with river-specific age-class structure
informed by bay-level parameters, and reef-specific age-class structure
informed by river-level structure. However, developing a viable initial
model for such a complex model hierarchy requires and in-depth
exploratory data analysis (EDA). The method in this paper is a tool that
facilitates comprehensive data visualizations for this objective. For
example, a BHM framework for these data would require a type of mixed
membership model (\emph{e.g.}, see Erosheva \& Fienberg
(\citeproc{ref-eroshevaBayesianMixedMembership2005}{2005})) that either
(a) regards each \(G_{Rrt}\) and \(G_{Rt}\) as an unknown parameter, or
(b) relies on fitting the full model hierarchy for different
combinations of \(G_{Rrt}\) and \(G_{Rt}\). The implementation of option
(a) is typically very complex and computationally intensive, unless
\(G_{Rrt}\) and \(G_{Rt}\) are constrained by highly informative priors
(\emph{e.g.}, see Stephens
(\citeproc{ref-stephensBayesianAnalysisMixture2000}{2000})) that may
result in unwanted bias. Option (b) requires the enumeration of all
plausible combinations of \(G_{Rrt}\) and \(G_{Rt}\), and the comparison
of model performance among the entire set of corresponding BHM fits, so
that the best performing set of \(G_{Rrt}\) and \(G_{Rt}\) values can be
identified. In other words, neither (a) nor (b) is practical without
some guidance based on an in-depth EDA. Our method can be regarded as
the EDA counterpart of option (b), thus, not only does it reduce the
computational burden that is required to fit a very large number of
complex BHMs, but it also provides data visualizations that shed light
on the ecological phenomena of age-class structure, lifespan, and
resilience, all in a holistic manner.

\subsection{Implications for
resiliency}\label{implications-for-resiliency}

\phantom{xxxx}Our research suggests that older age classes are becoming
more common throughout Virginia's Chesapeake Bay. Compared to studies in
the 2010s (\citeproc{ref-hardingManagementPiankatankRiver2010}{Harding
et al., 2010}; \citeproc{ref-mannPopulationStudiesNative2009}{Mann et
al., 2009};
\citeproc{ref-southworthOysterCrassostreaVirginica2010}{Southworth et
al., 2010}), we found that age three-and-older oysters have been present
more frequently, especially in the second decade (2014-2023). Between
1998 and 2009, nearly no oysters of age four were identified by Harding
et al. (\citeproc{ref-hardingManagementPiankatankRiver2010}{2010}). Our
study identified that age four-and-older oysters were present very
consistently after 2013. Before 2009, year-one and year-two age classes
in the James River had very high mortality rates despite consistently
strong recruitment (\citeproc{ref-mannPopulationStudiesNative2009}{Mann
et al., 2009}). Our study found consistent age groups from one to five,
although they were more similar in size (mostly less than market size of
76 mm). This suggests that our more rigorous method may be better
equipped to identify age groups in systems under stress compared to
previous methods applied (e.g.,
\citeproc{ref-bhattacharyaSimpleMethodResolution1967}{Bhattacharya,
1967};
\citeproc{ref-mildenbergerTropFishRPackageFisheries2017}{Mildenberger et
al., 2017}). Further, previous studies in Chesapeake Bay aggregated
oysters into 2-mm bin sizes
(\citeproc{ref-hardingManagementPiankatankRiver2010}{Harding et al.,
2010}; \citeproc{ref-mannPopulationStudiesNative2009}{Mann et al.,
2009};
\citeproc{ref-southworthOysterCrassostreaVirginica2010}{Southworth et
al., 2010}), which may have masked the small size differences between
ages two through four. We identified some age-six classes in reefs in
the James and Piankatank Rivers (see Appendix), suggesting that in terms
of age structure, the oyster reef population in Chesapeake Bay could be
improving due to resiliency\emph{,} especially to disease. This is
because the epizootics of \emph{Perkinsus marinus} and \emph{MSX} lead
to high mortality events, particularly of larger (older) oysters,
controlled by salinity and temperature
(\citeproc{ref-andrewsEpizootiologyDiseaseCaused1988}{Andrews, 1988};
\citeproc{ref-kusnerikUsingFossilRecord2018}{Kusnerik et al., 2018}).
Thus, the presence of older age groups suggests greater survival than
previously estimated by Harding et al.
(\citeproc{ref-hardingManagementPiankatankRiver2010}{2010}), Mann et al.
(\citeproc{ref-mannPopulationStudiesNative2009}{2009}); and Southworth
et al. (\citeproc{ref-southworthOysterCrassostreaVirginica2010}{2010}).

\phantom{xxxx}Our research found additional evidence of resiliency,
through the continuation of shell size shrinkage over time while
remaining alive. Many cohorts are less than market size in Virginia (76
mm), a similar finding to previous studies
(\citeproc{ref-mannOysterShellProduction2022}{Mann et al., 2022}). This
``shrinking'' trend is most apparent in the James River, which
historically has the largest acreage of productive bottom and highest
average annual production
(\citeproc{ref-havenPublicOysterBottoms1986}{Haven \& Whitcomb, 1986}).
The shrinking trend is least pronounced in the Piankatank and Great
Wicomico rivers, two river systems that have received shell plantings
since the 1960s, and designated sanctuary reefs through modern
large-scale restoration efforts
(\citeproc{ref-schulteHistoryVirginiaOyster2017}{Schulte, 2017}).
Nonetheless, oysters larger than 51 mm are very susceptible to mortality
from \emph{MSX}, which could explain smaller shell sizes throughout the
region in recent years
(\citeproc{ref-kusnerikUsingFossilRecord2018}{Kusnerik et al., 2018}).
Further, adult oysters display decreased growth rates and calcification
rates in response to ocean acidification
(\citeproc{ref-lemassonLinkingBiologicalImpacts2017}{Lemasson et al.,
2017}). These factors and their interactions with salinity and
temperature could be contributing to the signal of increased lifespan,
despite environmental decline.

\phantom{xxxx}There are two main hypotheses for explaining shell size
shrinkage over time. The first suggests human-induced shrinking:
fisheries science research has suggested that harvesting pressure on
bigger, faster growing animals has selected for slow-growing, smaller,
animals over time
(\citeproc{ref-allendorfHumaninducedEvolutionCaused2009}{Allendorf \&
Hard, 2009}; \citeproc{ref-rutterNaturalHistoryQuinnat1902}{Rutter,
1902}). The second suggests environmental-induced shrinking, such as
disease prevalence and coastal acidification could be drivers that slow
the growth and terminal size of oysters
(\citeproc{ref-kusnerikUsingFossilRecord2018}{Kusnerik et al., 2018};
\citeproc{ref-lemassonLinkingBiologicalImpacts2017}{Lemasson et al.,
2017}; \citeproc{ref-rossMetaanalysesRevealClimate2024}{Ross et al.,
2024}; \citeproc{ref-rybovichIncreasedTemperaturesCombined2016}{Rybovich
et al., 2016}). In particular, coastal and ocean acidification has the
potential to be an increasingly important contributor to oyster
challenges in Chesapeake Bay
(\citeproc{ref-boulaisOysterReproductionCompromised2017}{Boulais et al.,
2017}; \citeproc{ref-lemassonLinkingBiologicalImpacts2017}{Lemasson et
al., 2017};
\citeproc{ref-waldbusserOysterShellDissolution2011}{Waldbusser et al.,
2011}). These hypotheses have yet to be explored directly and
holistically in relation to resiliency for eastern oysters in the
Chesapeake Bay.

\phantom{xxxx}Our results suggest that eastern oysters are displaying
enhanced resiliency, due to increased longevity of cohorts, despite
continued environmental decline, exhibited by shell size shrinkage over
time. These smaller (yet older) age groups contribute less shell to the
reef, suggesting a complex push-and-pull in terms of self-sustainability
in the future. The current framework of ecological resiliency does not
aid resource management in these complex situations. Standard ecological
definitions of ``resiliency'', an inherently abstract concept, are
difficult to define. Baskett et al.
(\citeproc{ref-baskettResponseDiversityCan2014}{2014}) defines
``resiliency'' as the probability of maintaining a given state
(ecological resilience) and the rate of return to a given state
following a disturbance (engineering resilience). National Academies of
Science, Engineering, and Medicine
(\citeproc{ref-nationalacademiesofscienceengineeringandmedicineDecisionFrameworkInterventions2019}{2019})
defines resiliency the ability to survive despite climate change. These
definitions lack an explicit description of the metrics that should be
measured and the spatiotemporal scale at which to do so. Our goal is to
address the knowledge gap as we progress in our research.

\section{Conclusion}\label{conclusion}

\phantom{xxxx}Our paper is the first study on Virginia's Chesapeake Bay
oysters that estimates age groups and link them into age cohorts over an
expansive spatiotemporal scale. This method facilitated the
identification, tracking, and aging of oyster cohorts, and subsequently
revealing signals of resiliency. The results suggest that in the
post-epizootic period (post 2005) and period where ecosystem management
is stable (post 2009), age-truncation is being reversed, despite
continued shell size shrinkage in recent years. Mann \& Powell
(\citeproc{ref-mannWhyOysterRestoration2007}{2007}) call for novel and
alternative approaches that address oyster restoration holistically, and
our method provides an avenue to rigorously guide the formal modeling of
age structure and the identification of river systems that have not
rebuilt and may require a shift in management strategy. Our results
suggest that this type of modeling has the ptoentail to reveal that
differences in management strategies, despite continued environmental
decline, could be contributing to resiliency in Virginia's Chesapeake
Bay oyster reefs.

\section{Acknowledgements}\label{acknowledgements}

This research was supported by award \#24970 from the Chesapeake Bay
Trust-Chesapeake Oyster Alliance Chesapeake Oyster Innovation Award. The
authors acknowledge William \& Mary Research Computing for providing
computational resources and/or technical support that have contributed
to the results reported within this paper
(\url{https://www.wm.edu/offices/it/services/researchcomputing/atwm/})

\section{Data availability statement}\label{data-availability-statement}

The data, code, and interactive maps of the results are available at the
following GitHub repository: \url{https://go.wm.edu/GsmMyG}.

\section{\texorpdfstring{APPENDIX: results for remaining \(R\)
strata}{APPENDIX: results for remaining R strata}}\label{appendix-results-for-remaining-r-strata}

See results for remaining \(R\) strata at the following GitHub
repository: \url{https://go.wm.edu/GsmMyG}

\section*{References}\label{references}
\addcontentsline{toc}{section}{References}

\phantomsection\label{refs}


\begin{CSLReferences}{1}{0}
\bibitem[\citeproctext]{ref-allendorfHumaninducedEvolutionCaused2009}
Allendorf, F. W., \& Hard, J. J. (2009). Human-induced evolution caused
by unnatural selection through harvest of wild animals.
\emph{Proceedings of the National Academy of Sciences},
\emph{106}(supplement\_1), 9987--9994.
\url{https://doi.org/10.1073/pnas.0901069106}

\bibitem[\citeproctext]{ref-andrewsEpizootiologyDiseaseCaused1988}
Andrews, J. D. (1988). Epizootiology of the disease caused by the oyster
pathogen {Perkinsus} marinus and its effects on the oyster industry.
\emph{American Fisheries Society Special Publication}, \emph{18},
47--63.

\bibitem[\citeproctext]{ref-bairdSeasonalDynamicsChesapeake1989}
Baird, D., \& Ulanowicz, R. E. (1989). The seasonal dynamics of the
{Chesapeake Bay} ecosystem. \emph{Ecological Monographs}, \emph{59}(4),
329--364. \url{https://www.jstor.org/stable/1943071}

\bibitem[\citeproctext]{ref-baskettResponseDiversityCan2014}
Baskett, M. L., Fabina, N. S., \& Gross, K. (2014). Response {Diversity
Can Increase Ecological Resilience} to {Disturbance} in {Coral Reefs}.
\emph{The American Naturalist}, \emph{184}(2), E16--E31.
\url{https://doi.org/10.1086/676643}

\bibitem[\citeproctext]{ref-baylorSurveyOysterGrounds1895}
Baylor, J. (1895). Survey of oyster grounds in {Virginia} - report of
{J}.{B}. Baylor to the governor of {Virginia}. \emph{Miscellaneous}.

\bibitem[\citeproctext]{ref-bhattacharyaSimpleMethodResolution1967}
Bhattacharya, C. G. (1967). A simple method of resolution of a
distribution into {Gaussian} components. \emph{Biometrics},
\emph{23}(1), 115. \url{https://doi.org/10.2307/2528285}

\bibitem[\citeproctext]{ref-bonarControlOysterSettlement1990}
Bonar, D. B., Coon, S. L., Walch, M., Weiner, R. M., \& Fitt, W. (1990).
Control of {Oyster Settlement} and {Metamorphosis} by {Endogenous} and
{Exogenous Chemical Cues}. \emph{Bulletin of Marine Science},
\emph{46}(2), 484--498.

\bibitem[\citeproctext]{ref-boulaisOysterReproductionCompromised2017}
Boulais, M., Chenevert, K. J., Demey, A. T., Darrow, E. S., Robison, M.
R., Roberts, J. P., \& Volety, A. (2017). Oyster reproduction is
compromised by acidification experienced seasonally in coastal regions.
\emph{Scientific Reports}, \emph{7}(1), 13276.
\url{https://doi.org/10.1038/s41598-017-13480-3}

\bibitem[\citeproctext]{ref-coenRoleOysterReefs1999}
Coen, L., Luckenbach, M., \& Breitburg, D. (1999). The {Role} of {Oyster
Reefs} as {Essential Fish Habitat}: {A Review} of {Current Knowledge}
and {Some New Perspectives}. \emph{American Fisheries Society
Symposium}, \emph{22}, 438--454.

\bibitem[\citeproctext]{ref-commissionoffisheriesofvirginiaFiftyforthFiftyfifthAnnual1954}
Commission of Fisheries of Virginia. (1954). \emph{Fifty-forth and
fifty-fifth annual reports of the {Commission} of {Fisheries} of
{Virginia} (1954)} (p. 63).

\bibitem[\citeproctext]{ref-eroshevaBayesianMixedMembership2005}
Erosheva, E. A., \& Fienberg, S. E. (2005). Bayesian mixed membership
models for soft clustering and classification. In C. Weihs \& W. Gaul
(Eds.), \emph{Classification --- the {Ubiquitous Challenge}} (pp.
11--26). Springer. \url{https://doi.org/10.1007/3-540-28084-7_2}

\bibitem[\citeproctext]{ref-fabri-ruizBenthicEcoregionalizationBased2020}
Fabri-Ruiz, S., Danis, B., Navarro, N., Koubbi, P., Laffont, R., \&
Saucède, T. (2020). \emph{Benthic ecoregionalization based on echinoid
fauna of the {Southern Ocean} supports current proposals of {Antarctic
Marine Protected Areas} under {IPCC} scenarios of climate change}.
\url{https://doi.org/10.1111/gcb.14988}

\bibitem[\citeproctext]{ref-grabowskiEconomicValuationEcosystem2012}
Grabowski, J. H., Brumbaugh, R. D., Conrad, R. F., Keeler, A. G.,
Opaluch, J. J., Peterson, C. H., Piehler, M. F., Powers, S. P., \&
Smyth, A. R. (2012). Economic {Valuation} of {Ecosystem Services
Provided} by {Oyster Reefs}. \emph{BioScience}, \emph{62}(10), 900--909.
\url{https://doi.org/10.1525/bio.2012.62.10.10}

\bibitem[\citeproctext]{ref-haasEffectSpringneapTidal1977}
Haas, L. W. (1977). The effect of the spring-neap tidal cycle on the
vertical salinity structure of the {James}, {York} and {Rappahannock}
rivers, virginia, {U}.{S}.{A}. \emph{Estuarine and Coastal Marine
Science}, \emph{5}(4), 485--496.
\url{https://doi.org/10.1016/0302-3524(77)90096-2}

\bibitem[\citeproctext]{ref-hardingManagementPiankatankRiver2010}
Harding, J. M., Mann, R., Southworth, M. J., \& Wesson, J. A. (2010).
Management of the {Piankatank} river, {Virginia}, in support of oyster
({\emph{Crassostrea}}{ \emph{Virginica,}} {Gmelin} 1791) fishery
repletion. \emph{Journal of Shellfish Research}, \emph{29}(4), 867--888.
\url{https://doi.org/10.2983/035.029.0421}

\bibitem[\citeproctext]{ref-havenPublicOysterBottoms1986}
Haven, D. S., \& Whitcomb, J. P. (1986). \emph{The public oyster bottoms
in {Virginia}: An overview of their size, location, and productivity}.

\bibitem[\citeproctext]{ref-hemeonNovelShellStock2020}
Hemeon, K. M., Ashton-Alcox, K. A., Powell, E. N., Pace, S. M.,
Poussard, L. M., Solinger, L. K., \& Soniat, T. M. (2020). Novel shell
stock--recruitment models for {\emph{Crassostrea}}{ \emph{Virginica}} as
a function of regional shell effective surface area, a missing link for
sustainable management. \emph{Journal of Shellfish Research},
\emph{39}(3). \url{https://doi.org/10.2983/035.039.0310}

\bibitem[\citeproctext]{ref-kusnerikUsingFossilRecord2018}
Kusnerik, K. M., Lockwood, R., \& Grant, A. N. (2018). Using the fossil
record to establish a baseline and recommendations for oyster mitigation
in the {Mid-Atlantic U}.{S}. In C. L. Tyler \& C. L. Schneider (Eds.),
\emph{Marine {Conservation Paleobiology}} (pp. 75--103). Springer
International Publishing.
\url{https://doi.org/10.1007/978-3-319-73795-9_5}

\bibitem[\citeproctext]{ref-laslettFittingGrowthModels2004}
Laslett, G. M., Eveson, J. P., \& Polacheck, T. (2004). Fitting growth
models to length frequency data. \emph{ICES Journal of Marine Science},
\emph{61}(2), 218--230.
\url{https://doi.org/10.1016/j.icesjms.2003.12.006}

\bibitem[\citeproctext]{ref-lemassonLinkingBiologicalImpacts2017}
Lemasson, A. J., Fletcher, S., Hall-Spencer, J. M., \& Knights, A. M.
(2017). Linking the biological impacts of ocean acidification on oysters
to changes in ecosystem services: {A} review. \emph{Journal of
Experimental Marine Biology and Ecology}, \emph{492}, 49--62.
\url{https://doi.org/10.1016/j.jembe.2017.01.019}

\bibitem[\citeproctext]{ref-macdonaldAgeGroupsSizeFrequencyData1979}
Macdonald, P. D. M., \& Pitcher, T. J. (1979). Age-{Groups} from
{Size-Frequency Data}: {A Versatile} and {Efficient Method} of
{Analyzing Distribution Mixtures}. \emph{Journal of the Fisheries
Research Board of Canada}, \emph{36}(8), 987--1001.
\url{https://doi.org/10.1139/f79-137}

\bibitem[\citeproctext]{ref-mannWhyOysterRestoration2007}
Mann, R., \& Powell, E. N. (2007). Why oyster restoration goals in the
{Chesapeake Bay} are not and probably cannot be achieved. \emph{Journal
of Shellfish Research}, \emph{26}(4), 905--917.
\url{https://doi.org/10.2983/0730-8000(2007)26\%5B905:WORGIT\%5D2.0.CO;2}

\bibitem[\citeproctext]{ref-mannPopulationStudiesNative2009}
Mann, R., Southworth, M., Harding, J. M., \& Wesson, J. A. (2009).
Population studies of the native eastern oyster, {\emph{Crassostrea}}{
\emph{Virginica}}, ({Gmelin}, 1791) in the {James} river, {Virginia},
{USA}. \emph{Journal of Shellfish Research}, \emph{28}(2), 193--220.
\url{https://doi.org/10.2983/035.028.0203}

\bibitem[\citeproctext]{ref-mannOysterShellProduction2022}
Mann, R., Southworth, M., Wesson, J., Thomas, J., Tarnowski, M., \&
Homer, M. (2022). Oyster shell production and loss in the {Chesapeake
Bay}. \emph{Journal of Shellfish Research}, \emph{40}(3).
\url{https://doi.org/10.2983/035.040.0302}

\bibitem[\citeproctext]{ref-marquardtOysterAllometryGrowth2024}
Marquardt, A. R., Southworth, M., \& Mann, R. (2024). Oyster allometry:
Growth relationships vary across space. \emph{Journal of the Marine
Biological Association of the United Kingdom}, \emph{104}, e119.
\url{https://doi.org/10.1017/S0025315424001140}

\bibitem[\citeproctext]{ref-mildenbergerTropFishRPackageFisheries2017}
Mildenberger, T. K., Taylor, M. H., \& Wolff, M. (2017).
{\textbf{TropFishR}}: An {R} package for fisheries analysis with
length-frequency data. \emph{Methods in Ecology and Evolution},
\emph{8}(11), 1520--1527. \url{https://doi.org/10.1111/2041-210X.12791}

\bibitem[\citeproctext]{ref-nationalacademiesofscienceengineeringandmedicineDecisionFrameworkInterventions2019}
National Academies of Science, Engineering, and Medicine. (2019).
\emph{A decision framework for interventions to increase the persistence
and resilience of coral reefs} (p. 198). The National Academies Press.
\url{https://doi.org/10.17226/25424}

\bibitem[\citeproctext]{ref-neathBayesianInformationCriterion2012}
Neath, A. A., \& Cavanaugh, J. E. (2012). The {Bayesian} information
criterion: Background, derivation, and applications. \emph{WIREs
Computational Statistics}, \emph{4}(2), 199--203.
\url{https://doi.org/10.1002/wics.199}

\bibitem[\citeproctext]{ref-paceDyingDecayingDissolving2020}
Pace, S. M., Poussard, L. M., Powell, E. N., Ashton-Alcox, K. A.,
Kuykendall, K. M., Solinger, L. K., Hemeon, K. M., \& Soniat, T. M.
(2020). Dying, decaying, and dissolving into irrelevance: First direct
in-the-field estimate of {\emph{Crassostrea}}{ \emph{Virginica}}{
\emph{\emph{Shell Loss---a Case History from}} }{\emph{\emph{Mississippi
Sound}}}. \emph{Journal of Shellfish Research}, \emph{39}(2), 245.
\url{https://doi.org/10.2983/035.039.0206}

\bibitem[\citeproctext]{ref-powellRiseFallCrassostrea2012}
Powell, E. N., Klinck, J., Ashton-Alcox, K., Hofmann, E., \& Morson, J.
(2012). The rise and fall of {\emph{Crassostrea}}{ \emph{Virginica}}
oyster reefs: The role of disease and fishing in their demise and a
vignette on their management. \emph{Journal of Marine Research},
\emph{70}(2-3). \url{https://doi.org/10.1357/002224012802851878}

\bibitem[\citeproctext]{ref-powellHowLongDoes2006}
Powell, E. N., Kraeuter, J. N., \& Ashton-Alcox, K. A. (2006). How long
does oyster shell last on an oyster reef? \emph{Estuarine, Coastal and
Shelf Science}, \emph{69}(3-4), 531--542.
\url{https://doi.org/10.1016/j.ecss.2006.05.014}

\bibitem[\citeproctext]{ref-pritchardSalinityDistributionCirculation1952}
Pritchard, D. (1952). Salinity distribution and circulation in the
{Chesapeake Bay} estuarine system. \emph{Journal of Marine Research},
\emph{11}(2).

\bibitem[\citeproctext]{ref-rossMetaanalysesRevealClimate2024}
Ross, P. M., Pine, C., Scanes, E., Byrne, M., O'Connor, W. A., Gibbs,
M., \& Parker, L. M. (2024). Meta-analyses reveal climate change impacts
on an ecologically and economically significant oyster in {Australia}.
\emph{iScience}, \emph{27}(12).
\url{https://doi.org/10.1016/j.isci.2024.110673}

\bibitem[\citeproctext]{ref-rutterNaturalHistoryQuinnat1902}
Rutter, C. (1902). Natural history of the {Quinnat} salmon: A report of
investigations in the {Sacramento River} 1896-1901. \emph{Bulletin
United States Fish Commission}, \emph{22}, 65--141.

\bibitem[\citeproctext]{ref-rybovichIncreasedTemperaturesCombined2016}
Rybovich, M., Peyre, M. K. L., Hall, S. G., \& Peyre, J. F. L. (2016).
Increased {Temperatures Combined} with {Lowered Salinities
Differentially Impact Oyster Size Class Growth} and {Mortality}.
\emph{Journal of Shellfish Research}, \emph{35}(1), 101--113.
\url{https://doi.org/10.2983/035.035.0112}

\bibitem[\citeproctext]{ref-schulteHistoryVirginiaOyster2017}
Schulte, D. M. (2017). History of the {Virginia} oyster fishery,
{Chesapeake Bay}, {USA}. \emph{Frontiers in Marine Science}, \emph{4}.

\bibitem[\citeproctext]{ref-scruccaModelBasedClusteringClassification2023a}
Scrucca, L., Fraley, C., Murphy, T. B., \& Raftery, A. E. (2023).
\emph{Model-{Based Clustering}, {Classification}, and {Density
Estimation Using} mclust in {R}} (1st ed.). {Chapman and Hall/CRC}.
\url{https://doi.org/10.1201/9781003277965}

\bibitem[\citeproctext]{ref-sethiAccurateAgingJuvenile2017}
Sethi, S. A., Gerken, J., \& Ashline, J. (2017). Accurate aging of
juvenile salmonids using fork lengths. \emph{Fisheries Research},
\emph{185}, 161--168.
\url{https://doi.org/10.1016/j.fishres.2016.09.012}

\bibitem[\citeproctext]{ref-shawCohortAnalysisEuphausia2021}
Shaw, C. T., Bi, H., Feinberg, L. R., \& Peterson, W. T. (2021). Cohort
analysis of {\emph{Euphausia}}{ \emph{Pacifica}} from the {Northeast
Pacific} population using a {Gaussian} mixture model. \emph{Progress in
Oceanography}, \emph{191}, 102495.
\url{https://doi.org/10.1016/j.pocean.2020.102495}

\bibitem[\citeproctext]{ref-solingerOystersBegetShell2022}
Solinger, L. K., Ashton-Alcox, K. A., Powell, E. N., Hemeon, K. M.,
Pace, S. M., Soniat, T. M., \& Poussard, L. M. (2022). Oysters beget
shell and vice versa: Generating management goals for live oysters and
the associated reef to promote maximum sustainable yield of
{\emph{Crassostrea}}{ \emph{Virginica}}. \emph{Canadian Journal of
Fisheries and Aquatic Sciences}, \emph{79}(8), 1241--1254.
\url{https://doi.org/10.1139/cjfas-2021-0277}

\bibitem[\citeproctext]{ref-southworthOysterCrassostreaVirginica2010}
Southworth, M., Harding, J. M., Wesson, J. A., \& Mann, R. (2010).
Oyster ({\emph{Crassostrea}}{ \emph{Virginica}} , {Gmelin} 1791)
{Population Dynamics} on {Public Reefs} in the {Great Wicomico River},
{Virginia}, {USA}. \emph{Journal of Shellfish Research}, \emph{29}(2),
271--290. \url{https://doi.org/10.2983/035.029.0202}

\bibitem[\citeproctext]{ref-stephensBayesianAnalysisMixture2000}
Stephens, M. (2000). Bayesian {Analysis} of {Mixture Models} with an
{Unknown Number} of {Components- An Alternative} to {Reversible Jump
Methods}. \emph{The Annals of Statistics}, \emph{28}(1), 40--74.
\url{https://www.jstor.org/stable/2673981}

\bibitem[\citeproctext]{ref-vonbertalanffyQuantitativeTheoryOrganic1938}
von Bertalanffy, L. (1938). \emph{A quantitative theory of organic
growth (inquiries on growth laws {II})}. \emph{10}, 181--213.
\url{https://www.jstor.org/stable/41447359}

\bibitem[\citeproctext]{ref-waldbusserOysterShellDissolution2011}
Waldbusser, G. G., Steenson, R. A., \& Green, M. A. (2011). Oyster
{Shell Dissolution Rates} in {Estuarine Waters}: {Effects} of {pH} and
{Shell Legacy}. \emph{Journal of Shellfish Research}, \emph{30}(3),
659--669. \url{https://doi.org/10.2983/035.030.0308}

\bibitem[\citeproctext]{ref-weisePedigreeAnalysisEstimates2022}
Weise, E. M., Scribner, K. T., Adams, J. V., Boeberitz, O., Jubar, A.
K., Bravener, G., Johnson, N. S., \& Robinson, J. D. (2022). Pedigree
analysis and estimates of effective breeding size characterize sea
lamprey reproductive biology. \emph{Evolutionary Applications},
\emph{15}(3), 484--500. \url{https://doi.org/10.1111/eva.13364}

\bibitem[\citeproctext]{ref-zhouBayesianHierarchicalApproach2020}
Zhou, S., Martin, S., Fu, D., \& Sharma, R. (2020). A {Bayesian}
hierarchical approach to estimate growth parameters from length data of
narrow spread. \emph{ICES Journal of Marine Science}, \emph{77}(2),
613--623. \url{https://doi.org/10.1093/icesjms/fsz241}

\end{CSLReferences}
\end{document}